\documentclass[12pt]{article}
\usepackage[utf8]{inputenc}
\usepackage[T1]{fontenc}

\usepackage[nocompress]{cite}
\usepackage{IEEEtrantools,stackengine}
\stackMath

\usepackage{placeins}
\usepackage{authblk}
\usepackage{fullpage}
\usepackage{amssymb,amsmath}
\usepackage{float}
\usepackage{bm}
\usepackage{makecell}
\usepackage{siunitx}
\usepackage{csquotes}

\usepackage{xspace}
\newcommand\ie{i.e.\xspace}
\newcommand\eg{e.g.\xspace}

\newcommand\US{U.\,S.\xspace}

\def\sym#1{\ifmmode^{#1}\else\(^{#1}\)\fi}
\usepackage{booktabs}
\usepackage{longtable}
\usepackage{multirow}
\usepackage[export]{adjustbox}

\usepackage[dvipsnames]{xcolor}

\usepackage[left]{lineno}

\usepackage[
	colorinlistoftodos,
	textsize=footnotesize,
]{todonotes}

\definecolor{darkgreen}{rgb}{0.0, 0.5, 0.0}

\usepackage{setspace}
\usepackage[unicode=true]{hyperref}
\hypersetup{breaklinks=true,
	bookmarks=true,
	colorlinks=true,
	pdfborder={0 0 0},
	allcolors=black}
\expandafter\def\expandafter\UrlBreaks\expandafter{\UrlBreaks
	\do\-}

\usepackage[%
	sort&compress
]{cleveref}
\Crefname{appendix}{Supplement}{Supplements}

\usepackage{times}
\usepackage{enumerate}
\usepackage{float}

\usepackage{subcaption}
\usepackage{graphicx}

\graphicspath{{./Figures/}}

\usepackage[figuresright]{rotating}
\usepackage{tabularx}
\usepackage{dcolumn}
\usepackage{pdflscape}

\usepackage{array}

\newcolumntype{L}[1]{>{\raggedright\let\newline\\\arraybackslash\hspace{0pt}}p{#1}}
\newcolumntype{C}[1]{>{\centering\let\newline\\\arraybackslash\hspace{0pt}}p{#1}}
\newcolumntype{R}[1]{>{\raggedleft\let\newline\\\arraybackslash\hspace{0pt}}p{#1}}

\usepackage{sectsty}
\subsubsectionfont{\normalfont\itshape}

\makeatletter
\renewcommand{\fps@figure}{H}
\renewcommand{\fps@table}{H}
\makeatother

\allowdisplaybreaks

\usepackage{eurosym}

\usepackage{comment}

\usepackage{ntheorem}
\theoremseparator{:}

\usepackage[shortcuts]{extdash}
\usepackage{paralist}
\usepackage{colortbl} 
\usepackage{tabu}
\usepackage{bbm}
\usepackage{soul}

\DeclareSIUnit\week{week}
\DeclareSIUnit\weeks{weeks}
\DeclareSIUnit\billion{billion}

\begin{document}


\title{\centering\LARGE\singlespacing TikTok Rewards Divisive Political Messaging During the 2025 German Federal Election}

\renewcommand\Affilfont{\fontsize{9}{10.8}\selectfont}

\author[1]{Kirill Solovev}
\author[1]{Chiara Drolsbach}
\author[1]{Emma Demirel}
\author[1]{Nicolas Pröllochs
 \thanks{Correspondence: \url{nicolas.proellochs@wi.jlug.de}}}

\affil[1]{JLU Giessen, Germany}

\date{}

\maketitle


\clearpage
\begin{abstract}
	\normalfont
	\noindent
	Short-form video platforms like TikTok reshape how politicians communicate and have become important tools for electoral campaigning. Yet it remains unclear what kinds of political messages gain traction in these fast-paced, algorithmically curated environments, which are particularly popular among younger audiences. In this study, we use computational content analysis to analyze a comprehensive dataset of $N=$ \num{25292} TikTok videos posted by German politicians in the run-up to the 2025 German federal election. Our empirical analysis shows that videos expressing negative emotions (\eg, anger, disgust) and outgroup animosity were significantly more likely to generate engagement than those emphasizing positive emotion, relatability, or identity. Furthermore, ideologically extreme parties (on both sides of the political spectrum) were both more likely to post this type of content and more successful in generating engagement than centrist parties. Taken together, these findings suggest that TikTok's platform dynamics systematically reward divisive over unifying political communication, thereby potentially benefiting extreme actors more inclined to capitalize on this logic.
\end{abstract}

\flushbottom
\maketitle
\thispagestyle{empty}

\begin{center}
\begin{tabular}{p{14.5cm}}
\end{tabular}
\end{center}

\sloppy
\raggedbottom


\clearpage
\section*{Introduction}
\label{sec:introduction}


Short-form video platforms like TikTok have rapidly become central to electoral campaigning, particularly among younger audiences who consume political information in new formats \cite{Vazquez.2022,Newman.2022}. Unlike traditionally text-focused platforms such as Twitter/X or Facebook, TikTok emphasizes visual storytelling, short-form video, and trend-based interaction \cite{Zulli.2022}. Its algorithmically curated, highly personalized content feed allows political messages to circulate well beyond a politician's follower base, increasing the likelihood of visibility and engagement \cite{Zannettou.2024,Boeker.2022,Sangiorgio.2024}. For example, in Germany, TikTok counts almost \num{20.9} million users as of 2024 (\textasciitilde \SI{35}{\percent} of all adults \cite{BFDI.2024}), with a particularly high uptake among younger demographics (\textasciitilde \SI{55}{\percent} of users are under the age of 30 \cite{Statista.2024}). This makes TikTok a crucial channel for reaching first-time voters and politically less-engaged audiences who are less likely to consume traditional news media \cite{Cheng.2024,Vazquez.2022,Newman.2022}.


As platforms like TikTok gain political relevance, online political communication has become highly strategic. Politicians have strong incentives to actively craft content that shapes public narratives, mobilizes support, and maximizes engagement in a highly competitive attention economy \cite{Biswas.2025, Kreiss.2016}. To achieve this, politicians can draw on a wide repertoire of rhetorical tools \cite{Rathje.2025,Stieglitz.2012, Nulty.2016, Wollebaek.2019}. Examples include variation in sentiment (\eg, optimistic vs. pessimistic messages), the use of discrete emotions (\eg, fear to signal threat), and different forms of group-oriented messaging, which can either foster belonging through identity-based appeals \cite{Steffens.2013,Leach.2024}, or heighten division by targeting outgroups \cite{Nettasinghe.2025,Rathje.2021}. Such strategies may help politicians to resonate with specific voter segments, reinforce in-group loyalty, and signal alignment on salient issues \cite{Hackenburg.2023, Tornberg.2025}. Yet despite the variety of available strategies, it remains largely unclear which types of political content are actually rewarded by users and platform algorithms on short-form video platforms like TikTok. 


Prior research on online political communication has largely focused on primarily text-based social media platforms (\eg, X/Twitter, Facebook) and the political context of the \US For instance, affective content -- particularly negative sentiment and emotions such as anger -- has been shown to be associated with higher user engagement (\eg, shares, likes) \cite{Robertson.2023,Rathje.2021, Cheng.2024, Stieglitz.2013,Yu.2024}. Similarly, the use of outgroup language (\ie, expressing hostility toward political opponents) has been found to increase user engagement \cite{Nettasinghe.2025,Rathje.2021, Sangiorgio.2024,Yu.2024}. However, it is largely unclear to what extent these findings translate to short-form video platforms like TikTok, which operate under high levels of algorithmic curation \cite{Song.2024}, different demographics \cite{Gao.2023}, and audio-visual storytelling \cite{Cao.2025}. Despite TikTok’s growing role in political campaigning, especially among younger voters \cite{Medina.2020}, research on the platform is scant. The few existing studies in this direction 
have primarily focused on exploratory analyses of metadata (\eg, engagement or follower data \cite{Guinaudeau.2022,Ibrahim.2025,Li.2025}) or small-scale content analyses based on manually labeled samples \cite{Medina.2020,Zeng.2023,Cervi.2021,Zamora.2023,Moir.2023}. For instance, existing work suggests that political actors on TikTok frequently attempt to denigrate political opponents \cite{Zamora.2023} or foster partisanship \cite{Li.2025}. Yet, systematic empirical evidence on what content drives engagement in political communication on short-form video platforms is missing.


Here, we analyze a comprehensive dataset consisting of $N=$ \num{25292} TikTok videos posted by German politicians, political parties, and parliamentary groups in the run-up to the 2025 German federal election. This snap election followed a turbulent political period in Germany marked by the collapse of the previous center-left coalition in late 2024, and resulted in historic gains for far-right and far-left parties -- especially among young demographics \cite{Tagesschau.2025}. We implemented (and manually validated) a computational content analysis approach based on large language models (LLMs) to code each video for a wide range of content features, including \textit{Sentiment} (\ie, positive, neutral, or negative), discrete emotions (\eg, \textit{Anger}, \textit{Fear}), and group-oriented messaging such as \textit{Identity Language}, \textit{Relatability}, and \textit{Outgroup Animosity}. We then estimated multilevel regression models to assess which content characteristics were associated with higher levels of engagement on TikTok.


Our study contributes to research and ongoing debates about the electoral incentive dynamics in political communication on short-form video platforms. We demonstrate that videos featuring divisive rhetoric, especially those conveying \textit{Negative Sentiment}, \textit{Anger}, \textit{Disgust}, or \textit{Outgroup Animosity}, attract higher levels of engagement on TikTok. In contrast, content embedding \textit{Positive Sentiment}, \textit{Relatability}, or \textit{Identity Language} performs significantly worse. Furthermore, we find that extreme parties (on both sides of the political spectrum) were both more likely to post divisive content and more successful in generating engagement compared to their centrist counterparts. These observations suggest that TikTok's platform dynamics systematically reward divisive over unifying political communication, thereby benefiting political actors who are less constrained by norms of civility. Our findings have important implications for democratic discourse, the strategic incentives that shape political communication, and how young voters encounter politics in the video-first age.

\section*{Empirical Analysis}
\label{sec:results}

\subsection*{Posting Activity and Content Characteristics}

\begin{figure}
	\captionsetup[subfloat]{font={bf, small}, skip=0pt, singlelinecheck=false, labelformat=simple, position=top}
	\centering
	\includegraphics[width=0.8\textwidth]{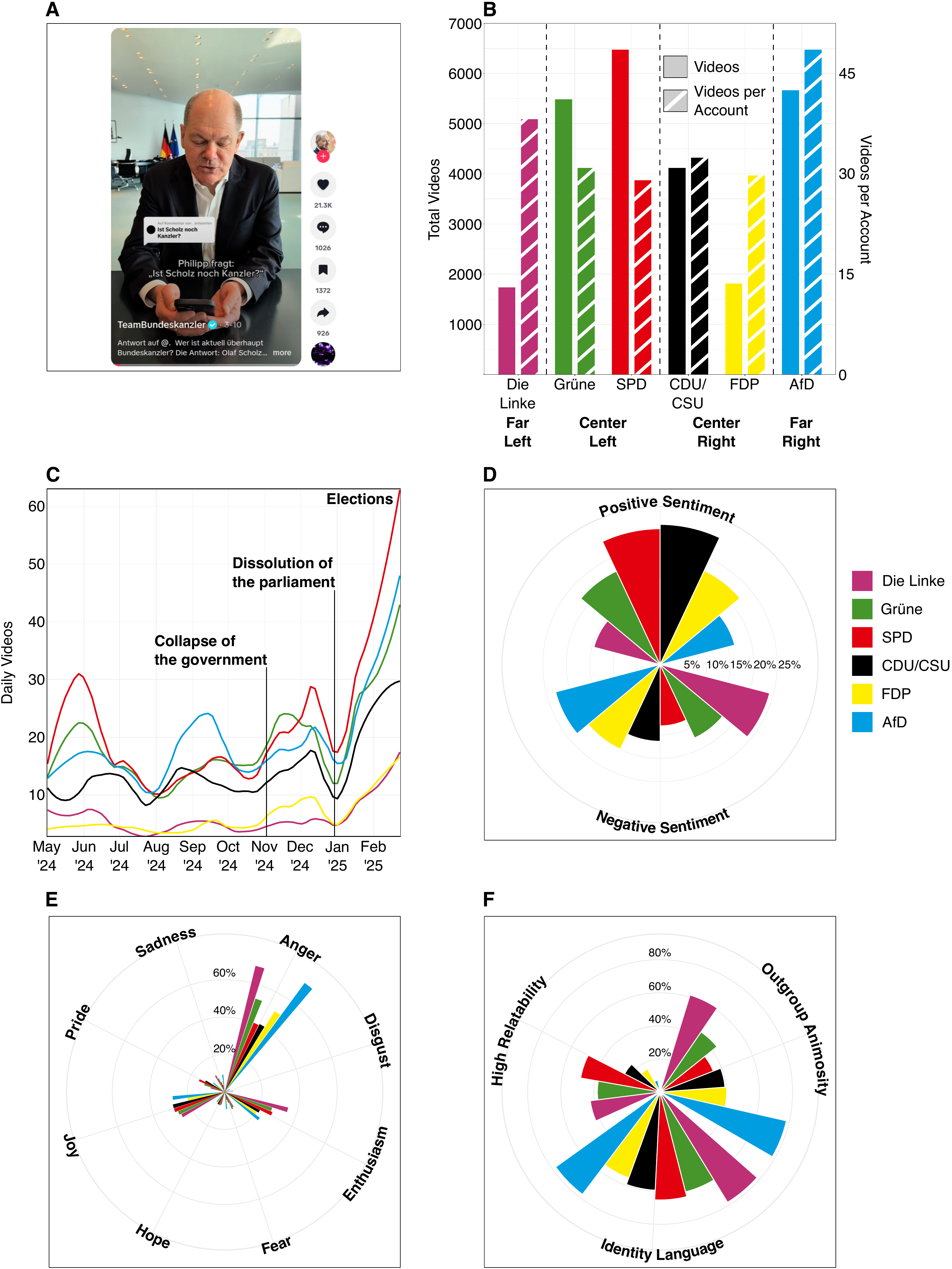}
	\caption{\textbf{Data overview.} \textbf{(A)}~An example of a campaign post on TikTok during the 2025 German federal election. \textbf{(B)}~Number of TikTok posts per political party and the average number of videos per account in our observation period from May 1, 2024 to February 23, 2025. \textbf{(C)}~The daily number of videos per party.  \textbf{(D)}~The share of posts (in \%) per  party expressing positive and negative sentiment. \textbf{(E)}~The share of posts (in \%) per party expressing discrete emotions. \textbf{(F)}~The share of posts (in \%) per party expressing outgroup animosity, identity language, and high relatability. Additional descriptive statistics  are in \Cref{supp:descriptives}.}
	\label{fig:data_overview}
\end{figure}

We used the TikTok Research API to collect all videos posted by accounts of German politicians (members of the national, state, or European parliaments), political parties, and parliamentary groups with an active TikTok account from May 1, 2024, until February 23, 2025, \ie during the nine months leading up to the German federal election (see Fig.~\ref{fig:data_overview}A for an example). In total, the dataset comprises \num{25292} videos from \num{727} distinct accounts, \SI{74.07}{\percent} of which were published by individual politicians (see \nameref{sec:methods}). The highest view counts for individual politicians were recorded for the accounts from then-Chancellor Olaf Scholz ($\approx 9.14$ million), Die-Linke lead candidate Heidi Reichinnek ($\approx 7.13$ million), and AfD lead candidate Alice Weidel ($\approx 6.82$ million). 

Fig.~\ref{fig:data_overview}B illustrates both the total volume of political video content (left axis) and the average number of videos per account (right axis) produced by German political parties between May 2024 and February 2025. The center-left \textit{SPD} (\ie, the incumbent party) published the highest volume of videos, totaling \num{6473}, followed by the far-right \textit{AfD} with \num{5665} videos. This surpassed the output from the center-right bloc, \ie, \textit{CDU/CSU} (\num{4118} videos) and \textit{FDP} (\num{1814} videos). \textit{Die Linke}, Germany’s primary far-left party, posted the fewest TikTok videos (\num{1734}). In terms of posting frequency per account, \textit{AfD} was the most active, averaging \num{50} videos per account. \textit{Die Linke} followed with an average of \num{40} videos across its \num{44} active accounts. Other parties averaged around \num{30} videos per account.

The video posting activity was closely tied to key political developments, with noticeable spikes following the collapse of the government in October 2024 and a surge after the dissolution of parliament in December 2024. Posting frequency also increased in the lead-up to several elections, including the European parliament election in June 2024, the regional election in three federal states (\ie, Brandenburg, Saxony, and Thuringia) in September 2024, and the federal election in February 2025, which marked the overall peak in activity (see Fig.~\ref{fig:data_overview}C). This temporal pattern suggests that political video production was highly responsive to institutional crises and the electoral timeline. 


Using a (manually validated) computational content analysis approach (see \nameref{sec:methods}), we coded each video for \textit{Sentiment} (\ie, positive, neutral, or negative), discrete emotions (\eg, \textit{Anger}, \textit{Fear}), and different forms of group-oriented messaging (\ie, \textit{Identity Language}, \textit{Relatability}, \textit{Outgroup Animosity}). Analysis of sentiment revealed systematic differences between political camps. Politicians from centrist parties (\ie, \textit{CDU/CSU, SPD, Grüne, FDP}) were more likely to produce positively valenced content than those at the extremes, namely \textit{Die Linke} and \textit{AfD} (see Figure~\ref{fig:data_overview}D). The proportion of positive videos was \SI{25}{\percent} for centrist, compared to around \SI{15}{\percent} for extreme parties. In comparison, less than \SI{20}{\percent} videos from centrist parties were negatively valenced, with the \textit{SPD} exhibiting the lowest proportion at \SI{13}{\percent}, suggesting a more optimistic messaging strategy. Extreme parties featured higher shares of negative sentiment. Specifically, \SI{24}{\percent} of videos from \textit{Die Linke} and \SI{23}{\percent} of videos from \textit{AFD} were negative, consistent with more critical or oppositional narratives.

An analysis of discrete emotions further underscored these differences. Across all parties, \textit{Anger} was the most dominant discrete emotion, but it appeared at roughly twice the rate in content from extreme parties (see Figure~\ref{fig:data_overview}E). The \textit{AfD} showed the highest overall prevalence, with over \SI{72}{\percent} of videos expressing \textit{Anger} (compared to \eg, \SI{40}{\percent} for the \textit{SPD}). In contrast, centrist parties conveyed more \textit{Hope} and \textit{Pride} than extreme parties, with relative frequencies of more than \SI{7}{\percent} for \textit{Hope} (vs. \SI{6}{\percent} for extreme parties) and \SI{11}{\percent} for \textit{Pride} (vs. \SI{7}{\percent} for extreme parties), thereby signaling a more prosocial emotional tone.

Group-oriented messaging also varied systematically across political parties (see Figure~\ref{fig:data_overview}F). Posts from extreme parties exhibited substantially higher levels of \textit{Outgroup Animosity} compared to more moderate parties. Specifically, approximately \SI{78}{\percent} of \textit{AfD} videos and \SI{61}{\percent} of \textit{Die Linke} videos contained outgroup animosity, which is substantially higher than the overall average of \SI{49}{\percent} across all parties.
Both the \textit{AfD} and \textit{Die Linke} also exhibited elevated use of \textit{Identity Language}, with both parties' shares being around \SI{78}{\percent}. Relatable content was most common among left-leaning parties, with \textit{SPD}, \textit{Grüne}, and \textit{Die Linke} posting relatable videos in \SI{48}{\percent}, \SI{37}{\percent}, and \SI{42}{\percent} of cases, respectively.

Taken together, these findings indicate clear affective and rhetorical distinctions between political camps: incumbent and center-left parties emphasized positive sentiment and relatable messaging, whereas extremist parties relied more heavily on emotional negativity, outgroup-focused rhetoric, and expressions of \textit{Anger}.

\subsection*{Drivers of Engagement}

\begin{figure}
	\captionsetup[subfloat]{font={bf, small}, skip=0pt, singlelinecheck=false, labelformat=simple, position=top}
	\centering
	\includegraphics[width=\textwidth]{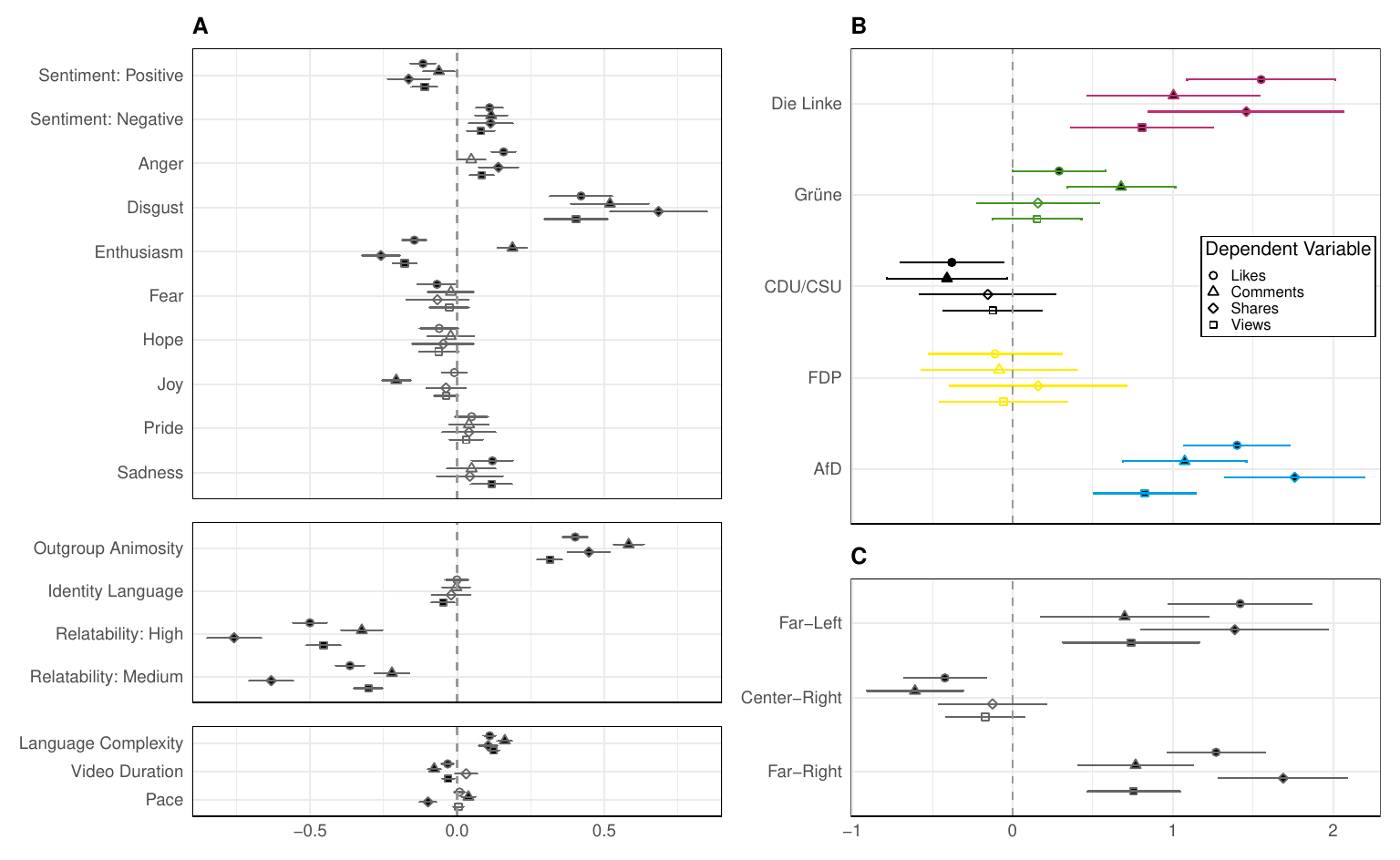}
	\caption{\textbf{Estimation results.} Multilevel negative binomial regression estimating the effects of \textbf{(A)} content characteristics (sentiment, discrete emotions, social identification), \textbf{(B)} political party, and \textbf{(C)} political alignment on the number of likes (circle), comments (triangle), shares (diamond), and views (square) on TikTok. Shown are the coefficient estimates with 95\% CIs. Account-specific random effects and monthly fixed effects are included. Full estimation results are in \Cref{supp:robustness}.}
	\label{fig:estimation_results}
\end{figure}

We estimated multilevel negative binomial regression models to examine how content features (\ie, sentiment, discrete emotions, and group-oriented messaging) were linked to engagement with videos on TikTok, as measured by the number of likes, comments, shares, and views (see \nameref{sec:methods}). The models control for the political orientation of the video’s author and structural features of the videos (language complexity, pace, and duration). To account for unobserved heterogeneity and temporal variation in platform dynamics, we additionally included account-specific random intercepts and month fixed effects. Figure \ref{fig:estimation_results} visualizes the coefficient estimates for the main explanatory variables; full estimation results are in SI, Supplement \ref{tbl:regression_dv}.


Across all outcome variables, videos expressing \textit{Negative Sentiment} were associated with significantly higher engagement. Relative to neutral content, they received $e^{0.110} - 1 \approx $\SI{11.6}{\percent} more likes (coef: $0.110$, $p < 0.001$), \SI{12.3}{\percent} more comments (coef: $0.116$, $p < 0.001$), \SI{12.1}{\percent} more shares (coef: $0.114$, $p < 0.001$), and \SI{8.3}{\percent} more views (coef: $0.080$, $p < 0.001$). In contrast, videos with \textit{Positive Sentiment} received \SI{10.9}{\percent} fewer likes (coef: $-0.116$, $p < 0.001$), \SI{6.0}{\percent} fewer comments (coef: $-0.062$, $p < 0.001$), \SI{15.2}{\percent} fewer shares (coef: $-0.165$, $p < 0.001$), and \SI{10.5}{\percent} fewer views (coef: $-0.111$, $p < 0.001$). Overall, this implies that negatively valenced political content was more likely to attract engagement on TikTok during the 2025 German federal election than either neutral or positively valenced content.

The results for discrete emotions mirror those observed for negative vs. positive sentiment. For the group of negatively valenced emotions, the largest effect size was observed for \textit{Disgust}, which was linked to \SI{52.3}{\percent} more likes (coef: $0.421$, $p < 0.001$), \SI{68.0}{\percent} more comments (coef: $0.519$, $p < 0.001$), \SI{98.4}{\percent} more shares (coef: $0.685$, $p < 0.001$), and \SI{49.8}{\percent} more views (coef: $-0.404$, $p < 0.001$). Similarly, the presence of \textit{Anger} predicted \SI{17.1}{\percent} more likes (coef: $0.158$, $p < 0.001$), \SI{15.0}{\percent} more shares (coef: $0.140$, $p < 0.001$), and \SI{8.8}{\percent} more views (coef: $0.084$, $p < 0.001$), but showed no significant effect for comments ($p > 0.05$). \textit{Sadness} was also associated with \SI{12.8}{\percent} more likes (coef: $0.120$, $p < 0.001$) and \SI{12.5}{\percent} more views (coef: $0.118$, $p < 0.01$), but showed no significant effects for comments ($p > 0.05$) or shares ($p > 0.05$). \textit{Fear} was linked to \SI{6.7}{\percent} fewer likes (coef: $-0.069$, $p < 0.05$), but had no significant effect on other engagement measures. In contrast, positively valenced emotions were largely associated with lower engagement. \textit{Joy} significantly decreased the number of comments a video received by \SI{18.7}{\percent} (coef: $-0.207$, $p < 0.001$), and showed no significant reductions across other metrics (each $p > 0.05$). \textit{Enthusiasm} displayed a mixed pattern: while it was associated with \SI{20.8}{\percent} more comments (coef: $0.189$, $p < 0.001$), it corresponded to lower values across the remaining outcomes, including \SI{13.5}{\percent} fewer likes (coef: $-0.145$, $p < 0.001$), \SI{22.8}{\percent} fewer shares (coef: $-0.259$, $p < 0.001$), and \SI{16.3}{\percent} fewer views (coef: $-0.178$, $p < 0.001$). No statistically significant effects were observed for  \textit{Hope}, and \textit{Pride} across any engagement metric (each $p > 0.05$). Taken together, these findings suggest that negative emotional appeals are more likely to elicit engagement on TikTok than positive ones.

The presence of group-oriented messaging was also associated with differences in engagement outcomes. Compared to videos with \textit{Low Relatability}, videos with \textit{Medium Relatability} received \SI{30.6}{\percent} fewer likes (coef: $-0.365$, $p < 0.001$), \SI{19.9}{\percent} fewer comments (coef: $-0.222$, $p < 0.001$), \SI{46.8}{\percent} fewer shares (coef: $-0.632$, $p < 0.001$), and \SI{26.0}{\percent} fewer views (coef: $-0.302$, $p < 0.001$). \textit{High Relatability} showed even larger reductions, including \SI{39.4}{\percent} fewer likes (coef: $-0.501$, $p < 0.001$), \SI{27.8}{\percent} fewer comments (coef: $-0.326$, $p < 0.001$), \SI{53.2}{\percent} fewer shares (coef: $-0.760$, $p < 0.001$), and \SI{36.5}{\percent} fewer views (coef: $-0.455$, $p < 0.001$). \textit{Identity Language} (\eg, references to shared group membership) was significantly associated with \SI{4.7}{\percent} fewer views (coef: $-0.048$, $p < 0.05$), but showed no significant effect on any other engagement metric (each $p>0.05$). In contrast, \textit{Outgroup Animosity} (\eg, expressions of hostility or negative attitudes toward other groups) was associated with \SI{49.3}{\percent} more likes (coef: $0.401$, $p < 0.001$), \SI{79.2}{\percent} more comments (coef: $0.583$, $p < 0.001$), \SI{56.6}{\percent} more shares (coef: $0.448$, $p < 0.001$), and \SI{37.2}{\percent} more views (coef: $0.316$, $p < 0.001$). Overall, these findings imply that content emphasizing group division was more likely to elicit engagement than content appealing to belonging (\ie, \textit{Relatability} and \textit{Identity Language}) during the 2025 German federal election.


Beyond content characteristics, political affiliation was also strongly associated with variation in engagement. Politicians from the two ideologically extreme parties in Germany, \textit{Die Linke} and \textit{AfD}, consistently outperformed those from centrist parties. On average, compared to videos posted by \textit{SPD} politicians, videos from \textit{Die Linke} politicians received \SI{371.6}{\percent} more likes (coef: $1.551$, $p<0.001$), \SI{172.9}{\percent} more comment (coef: $1.004$, $p<0.001$), \SI{329.4}{\percent} more shares (coef: $1.459$, $p<0.001$), and \SI{124.5}{\percent} more views (coef: $0.809$, $p<0.001$). Similarly, videos from \textit{AfD} politicians received \SI{206.0}{\percent} more likes (coef: $1.401$, $p<0.001$), \SI{192.7}{\percent} more comments (coef: $1.074$, $p<0.001$), \SI{482.0}{\percent} more shares (coef: $1.761$, $p<0.001$), and \SI{128.2}{\percent} more views (coef: $0.825$, $p<0.001$). In contrast, the observed differences among centrist parties (\ie, \textit{SPD, CDU/CSU, FDP, Grüne}) were rather small and largely not statistically significant (see \Cref{fig:estimation_results}B). These patterns were further confirmed when controlling for ideological alignment instead of party affiliation (see \nameref{sec:methods}). Also with this model specification, videos from both \textit{Far-Left} and \textit{Far-Right} accounts received significantly more likes, comments, shares, and views compared to videos from \textit{Center-Left} oriented accounts (each $p < 0.001$; see \Cref{fig:estimation_results}C).


Our models also account for potentially confounding variables, namely, \textit{Language Complexity}, \textit{Duration}, and \textit{Pace}. We find that higher \textit{Language Complexity} was associated with more likes, comments, shares, and views (each $p < 0.001$). In contrast, longer \textit{Duration} was linked to fewer likes  ($p<0.01$) and views ($p<0.01$), as well as fewer comments ($p < 0.001$). The \textit{Pace} of a video shows a negative and significant effect on comments ($p < 0.01$) and shares (both $p < 0.001$), but not on likes or views. Notably, however, the effect sizes for these control variables are comparatively small (see \Cref{fig:estimation_results}). 

\label{supp:estimation_results}

Multiple exploratory analyses extended our results and confirmed their robustness (see \Cref{supp:robustness}). First, we assessed multicollinearity by calculating the variance inflation factor (VIF) for all explanatory variables separately for each outcome and political affiliation variable. All VIFs remained below the critical threshold of four (see \Cref{tbl:vif}). Second, we estimated an implausible model using the number of \textit{XYZ} in each transcript as an explanatory variable \cite{Feuerriegel.2025}. Across all outcome variables, implausible models resulted in higher AIC values (\ie, lower model adequacy) and effect sizes close to zero (see \Cref{tbl:regression_xyz}). Third, we repeated our analysis with continuous measures for sentiment and emotion, which yielded consistent results (see \Cref{tbl:regression_continuous}). Fourth, we implemented a zero-inflated beta regression using the number of likes, comments, and shares per view as the dependent variables (see \Cref{tbl:regression_alternative}), thereby distinguishing active engagement (\ie, likes, comments, and shares) from passive engagement (\ie, views without likes, comments, or shares). Here, we found that especially moralized language such as \textit{Outgroup Animosity}, \textit{Identity Language}, \textit{Anger}, and \textit{Disgust} were among the strongest predictors of active engagement. In contrast, low arousal emotions such as \textit{Hope} and \textit{Sadness} were either negatively associated or not statistically significant. 
Altogether, our exploratory analyses provided confirmatory evidence that negative and divisive messaging was more effective than appeals to belonging in driving engagement on TikTok during the 2025 German federal election.

\newpage

\section*{Discussion}
\label{sec:discussion}

Here, we analyzed a comprehensive dataset consisting of $N=$ \num{25292} TikTok videos posted by German politicians, political parties, and parliamentary groups in the run-up to the 2025 German federal election. We found that videos featuring divisive rhetoric, especially those conveying \textit{Negative Sentiment}, \textit{Anger}, \textit{Disgust}, or \textit{Outgroup Animosity}, attract higher levels of engagement on TikTok. In contrast, content emphasizing positive emotion through \textit{Positive Sentiment}, \textit{Identity Language}, or \textit{Relatability} performs significantly worse. Moreover, we observed that extreme parties (on both sides of the political spectrum) were both more likely to post divisive content and more successful in generating engagement compared to their centrist counterparts. Taken together, these findings shed light on the strategic incentives that shape political communication in the video-first era and suggest that TikTok's platform dynamics systematically reward divisive over unifying content, thereby potentially advantaging extreme actors who are more inclined to adopt such messaging strategies. 

Prior research on online political communication has largely focused on primarily text-based social media platforms (\eg, X/Twitter, Facebook) and the political context of the \US \cite{Rathje.2021,Brady.2017,Falkenberg.2024,Guess.2023,Dang.2017,Bossetta.2018}. Concordant with the majority of these works, we find that negative and divisive content was particularly likely to go viral on TikTok during the 2025 German federal election. Our results also reinforce prior evidence \cite{Rathje.2021} on the virality of moral emotions (\eg, \textit{Anger}, \textit{Disgust}, and \textit{Outgroup Animosity}) on social media -- which showed the strongest positive effects on engagement in our dataset. At the same time, our results challenge common notions often drawn from adjacent fields such as influencer marketing: we find that content emphasizing \textit{Relatability} -- typically considered a key driver of engagement \cite{Atiq.2022,Liu.2024} -- was associated with the strongest negative effect on engagement. A plausible explanation may lie in perceived authenticity. While audiences tend to relate to influencers who appear similar to themselves, politicians are often seen as strategic actors \cite{Aalberg.2012}, which may make their attempts at relatability seem insincere or performative rather than authentic \cite{Enli.2015}. This highlights the challenges political actors face on short-form video platforms when attempting to align with the expectations of younger demographics.

Our findings point towards considerable incentive problems for political actors. When platforms and their algorithms reward divisive content with heightened visibility and engagement, politicians may feel pressured to adopt more extreme, provocative, or polarizing rhetoric. Prior research suggests that such content can have serious negative implications for societies, for example, by fostering anti-democratic attitudes, deepening partisan divides, and reinforcing discrimination against marginalized groups \cite{Hartman.2022,Bail.2018, Garrett.2020,Vaidhyanathan.2018,Lorenz.2023}.  At the same time, unifying communication strategies (\eg, positive sentiment, relatable narratives) are disincentivized. Concerningly, this incentive structure, favoring divisive over unifying content, creates a potential asymmetric advantage for extremist parties, which are often more willing to exploit this logic. In contrast, moderate parties, which are typically more constrained by norms of civility and democratic responsibility \cite{Gidron.2020}, face greater difficulty achieving comparable visibility and engagement. This asymmetry poses the risk of further accelerating political radicalization in digital societies. 


As with all research, ours is not free of limitations and offers opportunities for future research. First, our findings are based on observational data and we therefore refrain from making causal claims. Future research could use (quasi-)experimental designs to disentangle algorithmic effects from genuine user preferences, though such efforts may be constrained by limited data access. Second, our analysis focuses on a single short-form video platform (\ie, TikTok) and the political context of Germany. While this complements earlier work \cite{Rathje.2021,Brady.2017,Falkenberg.2024,Guess.2023,Dang.2017,Bossetta.2018} centered on text-based platforms (\eg, Twitter/X, Facebook) and the \US political landscape, further research is needed to examine how our findings translate to other short-form video platforms and political contexts (\eg, in the Global South \cite{Falkenberg.2024,Munzert.2025}). Third, our content analysis depends on the accuracy of LLM-based annotations of content characteristics. However, we conducted extensive manual validation and prior research \cite{Feuerriegel.2025,Rathje.2024} implies that our approach is an effective tool for psychological content analysis that significantly outperforms methods that have traditionally been used (\eg, dictionary-based approaches). Fourth, our analysis focused on video transcripts and did not include visual elements (\eg, visual sentiment cues). While incorporating such data could potentially enrich the analysis, recent research suggests that text-based approaches often yield similar conclusions, with limited additional insight gained from visual data \cite{Chen.2025}. Fifth, more research is needed to identify effective interventions that improve political discourse on platforms like TikTok. Potential approaches include media literacy initiatives, greater algorithmic transparency, and platform design changes aimed at reducing the spread of divisive content \cite{Milli.2025,Combs.2023}. Here, a key goal should be to prevent a self-reinforcing cycle, in which even well-meaning political actors feel pressured to adopt more divisive communication strategies. 

From a broader perspective, our study raises urgent questions about the growing political influence of short-form video platforms and their algorithmic recommendation systems. Social media, once envisioned as a tool to connect individuals and democratize communication, has evolved into a channel where divisive content gains disproportionate visibility and engagement. This is especially concerning given that platforms like TikTok attract a disproportionately young user base -- a key voting demographic that may be particularly vulnerable to this kind of strategically framed political messaging. Ensuring this group has access to accurate and trustworthy information is essential for safeguarding electoral integrity. While we do not claim a causal link between success on TikTok and electoral outcomes, it is striking that the parties most successful on the platform were also the most favored by young voters in the 2025 German federal election \cite{Statista.2025}. 


\clearpage
\section*{Methods}
\label{sec:methods}

\subsection*{Data Collection}
\label{sec:data}

To systematically identify relevant TikTok accounts, we used a precompiled dataset monitoring the social media activities of German politicians and parties \cite{Fuchs.2025}. We filtered the list to include all active TikTok accounts belonging to German politicians who held office during our observation period in the national parliament, state parliaments, or the European Parliament. Additionally, we included active accounts of major political parties, defined as those represented in the most recent German national parliament (\textit{\ie, CDU/CSU, SPD, Die Grüne, Die Linke, FDP, AfD}). The final dataset included a total of \num{727} active TikTok accounts. To ensure accuracy and validity, we manually verified a random sample of 10\% of the accounts and confirmed that all belonged to the indicated political figures. 

Using the TikTok Research API \cite{TikTok.nn}, we then collected all videos posted by these accounts in the period from May 1, 2024, until the federal election day on February 23, 2025. To avoid capturing post-election content, we ended data collection at 6 pm on election day, when polls officially closed. For each video, we retrieved video metadata, including the video IDs, video length, username, hashtags, video description, automatic transcriptions, as well as engagement metrics, namely, the number of likes, comments, shares, and views. This procedure yielded a final collection of \num{25292} TikTok videos. We downloaded the corresponding video files using a custom crawler script based on \texttt{yt-dlp} \cite{yt-dlp.2025}.

To analyze the video content, we used \texttt{Whisper}, an open-source automatic speech recognition system, to transcribe the audio content of the video files \cite{OpenAi.2022,Radford.2022}. The resulting transcripts were then used to extract various content features (see next section). In addition, we measured three key structural characteristics of each video: (i) video \textit{Duration} (in seconds), (ii) \textit{Language Complexity}, measured using the Gunning-Fog index \cite{Gunning.1952,Solovev.2025}, and (iii) speaking \textit{Pace}, calculated as the number of words divided by the video's duration \cite{Sihag.2023,Welbourne.2016}. 

\subsection*{Content Annotation}
The video transcripts served as the basis for extracting various content features. Specifically, we annotated each transcript for the \textit{Sentiment} \cite{Feuerriegel.2025,Robertson.2023}, discrete \textit{Emotions} \cite{Feuerriegel.2025, Robertson.2023}, and three group-oriented messaging variables: \textit{Identity Language} \cite{Hopkins.2025,Rasmussen.2025}, \textit{Outgroup Animosity} \cite{Brady.2017,Yu.2024}, and \textit{Relatability} \cite{Atiq.2022,Liu.2024}. All annotation procedures followed best practices in natural language processing for behavioral science \cite{Feuerriegel.2025}. 

\textbf{Sentiment:}
We used the model \textit{tabularisai/multilingual-sentiment-analysis} to classify each sentence in the transcripts into one of five sentiment categories: 'very negative' (assigned a score of $-2$), 'negative' ($-1$), 'neutral' ($0$), 'positive' ($1$) or 'very positive' ($2$). To calculate a video-level sentiment score, we computed the arithmetic mean of the sentence-level scores. For better interpretability and to enable comparison with other categorical content features, we transformed the resulting scores into a categorical variable. Specifically, we coded videos as 'negative' (mean below $-0.5$), 'neutral' (mean between $-0.5$ and $0.5$), or 'positive' (mean greater than $0.5$). As a robustness check, we also repeated our analysis with continuous sentiment scores. Here, we found consistent results (see SI, \Cref{supp:robustness}).

\textbf{Discrete emotions:}
For emotion extraction, we employed an updated version of the model by \cite{Widmann.2022}, namely \textit{tweedmann/pol\_emo\_mDeBERTa2}. This model is fine-tuned to predict the probability of eight discrete emotions: \textit{Anger}, \textit{Disgust}, \textit{Enthusiasm}, \textit{Fear}, \textit{Hope}, \textit{Joy}, \textit{Pride}, and \textit{Sadness} in political German-language text. Each sentence was assigned a probability for each emotion, and emotions were considered present if their probability exceeded a threshold of 0.65 \cite{Widmann.2022}. At the video level, an emotion was considered present (=$1)$ if it appeared in at least one sentence (otherwise $=0$). We also tested an alternative aggregation approach (\ie, calculating mean emotion scores across sentences). Here, we found consistent results (see SI, \Cref{supp:robustness}).

\textbf{Group-oriented messaging:}
To identify the presence of \textit{Identity Language}, \textit{Outgroup Animosity}, and \textit{Relatability} in the videos, we employed a large language model (LLM)-based annotation approach. Specifically, we deployed an OpenAI Assistant (based on \textit{gpt-4-turbo}) configured to act as a “professional annotator specializing in political discourse analysis.” To guide the annotation, the assistant was provided with detailed definitions of the three variables. The variables identity language and outgroup animosity were coded as binary (\ie, T/F), while relatability was coded as a three-level categorical variable (\ie, low, medium, high). A full version of the annotation prompt is provided in the SI, \Cref{supp:llm_annotation}.

\textbf{User study:} We conducted a user study on Prolific to validate the automated identification of content features (see SI, \Cref{supp:validation} for details). Twenty native-speaking participants were recruited and presented with a randomly selected sample of 100 videos. For each video, they were asked to indicate whether the video contained any of the content features annotated by the LLM.
All participants had at least a bachelor's degree and reported German as their first language ($9$ females and $11$ males with a mean age of $39.55$ years). Participants were \num{1.8} times to \num{7.7} times more likely to report a content feature as present when the LLM had labeled the video as containing it, compared to when it had labeled it as not containing it. Across all content characteristics, the differences were statistically significant (all $p < 0.05$) and accompanied by moderate to high agreement among raters (ICC ranging from $0.345$ to $0.895$). 
This validates the annotation process and confirms that it captures content features in a way that aligns with human assessments. 

\subsection*{Regression Analysis}
We implemented hierarchical regression models with user-specific random effects to estimate how content characteristics are linked to user engagement with political videos on TikTok. This modeling approach allows us to account for both the fixed effects of content-level variables and unobserved heterogeneity across user accounts (\eg, varying social influence).

In our analysis, we focus on four engagement outcomes (\ie, dependent variables): the number of (i) likes, (ii) comments, (iii) shares, and (iv) views. Given that the variance of each of these engagement metrics is higher than their mean, an adjustment for overdispersion is necessary. Following best practices \cite{Hardin.2007}, we thus employ negative binomial regression models to explain the four engagement measures. Each outcome variable was modeled separately. For each, we estimate two variants: one that includes political party affiliation and the other that includes ideological alignment (\ie, left vs. right) as predictors, resulting in a total of eight model specifications.

Our primary explanatory variables are the content features extracted from the video transcripts. These include the \textit{Sentiment} (\ie, Negative, Neutral [reference], Positive), binary indicators for eight discrete emotions (\ie, \textit{Anger}, \textit{Disgust}, \textit{Enthusiasm}, \textit{Fear}, \textit{Hope}, \textit{Joy}, \textit{Pride}, \textit{Sadness}), and three variables capturing group-oriented messaging: \textit{Outgroup Animosity} ($= 1$ if present, otherwise $0$), \textit{Identity Language} ($= 1$ if present, otherwise $0$), and \textit{Relatability} (\ie, Low [reference], Medium, High). 

We further controlled for several structural features of the videos, namely, \textit{Language Complexity} (measured using the Gunning-Fog index), speaking \textit{Pace} (measured in words per second), and video \textit{Duration} (in seconds). To account for temporal variation (\eg, growth of follower bases and differences in account age), we included monthly fixed effects (\ie, $\mathit{Month}$). All continuous predictors were $z$-standardized to facilitate interpretability. 

Based on these variables, we specified the following regression model:
\begin{equation}
\begin{aligned}
    \log(E(\text{Y}_{ij}|\text{X}_{ij})) = & \, \beta_0 + \beta_1 \:\text{Sentiment}_{ij} + \beta_2 \: \text{Anger}_{ij} + \beta_3 \: \text{Disgust}_{ij} + \beta_4 \: \text{Enthusiasm}_{ij} \\&+ \beta_5 \: \text{Fear}_{ij} + \beta_6 \: \text{Hope}_{ij} + \beta_7 \: \text{Joy}_{ij}  + \beta_8 \: \text{Pride}_{ij} + \beta_9 \: \text{Sadness}_{ij} \\& + \beta_{10} \: \text{OutgroupAnimosity}_{ij} + \beta_{11} \: \text{IdentityLanguage}_{ij} + \beta_{12} \: \text{Relatability}_{ij} \\& + \beta_{13} \: \text{PoliticalOrientation}_{ij}  + \beta_{14} \: \text{LanguageComplexity}_{ij} + \beta_{15} \: \text{Pace}_{ij} \\&  + \beta_{16} \: \text{Duration}_{ij} + \beta_{17} \: \text{Month}_{ij} + \text{u}_{ij}.
\label{eq:model_spec}
\end{aligned}
\end{equation}

In Equation \ref{eq:model_spec}, $Y_{ij}$ denotes the outcome (\ie, the number of likes, comments, shares, or views) for video $i$ from account $j$, $\mathit{X}_{ij}$ is the vector of explanatory variables, $\beta_0$ represents the intercept, and $u_j$ are the account-specific random effects. The term $\mathit{PoliticalOrientation}_{ij}$ is a placeholder for either political party affiliation or ideological alignment, which are included in separate model specifications.
In the political party models, \textit{SPD} (\ie, the incumbent party) is used as the reference category, with additional categories for the other parties (\ie, \textit{Die Linke}, \textit{Grüne}, \textit{CDU/CSU}, \textit{FDP}, and \textit{AfD}). In the ideological alignment models, \textit{Center-Left} serves as the reference, with \textit{Far-Left}, \textit{Center-Right}, and \textit{Far-Right} as additional categories. 

We estimated all models using maximum likelihood estimation with the \texttt{glmmTMB} package in R 4.3.0.











\clearpage
\bibliography{literature.bib}


\clearpage
\appendix

\renewcommand{\thetable}{S\arabic{table}}
\renewcommand{\thefigure}{S\arabic{figure}}
\setcounter{figure}{0}
\setcounter{table}{0}

\begin{minipage}[t]{\textwidth}
	\nolinenumbers
	\begin{center}
		\huge\bfseries Supplementary Materials
	\end{center}
	\vspace{1cm}
\end{minipage}

\tableofcontents

\newpage
\section{Descriptive Statistics}
\label{supp:descriptives}

Summary statistics of our dataset are in \Cref{tbl:descriptives}.

\begin{table}[H]
	\caption{Descriptive Statistics.}
	\centering
	{
		\sisetup{table-space-text-post = {}}
	\scriptsize
	\begin{tabular}{lS[table-format=4.3]S[table-format=4.3]S[table-format=4.3]S[table-format=4.3]S[table-format=4.3]}

		\toprule
		\textbf{Variable}    & \multicolumn{1}{c}{\textbf{Mean}} & \multicolumn{1}{c}{\textbf{Median}} & \multicolumn{1}{c}{\textbf{Min}} & \multicolumn{1}{c}{\textbf{Max}} & \multicolumn{1}{c}{\textbf{SD}} \\
		\midrule
		\multicolumn{6}{c}{\textsc{Dependent variables}}                                                                                                                                                       \\
		\midrule
		Like Count (in 1000)      & 2.891 & 0.104 & 0.000 & 868.868 & 19.559 \\
		Comment Count (in 1000)   & 0.192 & 0.012 & 0.000 & 47.798 & 0.868 \\
        Share Count (in 1000)     & 0.252 & 0.003 & 0.000 & 260.202 & 2.752 \\ 
		View Count (in 1000)      & 37.616 & 1.509 & 0.000 & 11863.196 & 226.949 \\
		\midrule
		\multicolumn{6}{c}{\textsc{Independent variables}}                                                                                                                                                     \\
		\midrule
		Sentiment: Positive       & 0.233 & 0.000 & 0.000 & 1.000 & 0.423 \\
		Sentiment: Negative       & 0.180 & 0.000 & 0.000 & 1.000 & 0.384 \\
		Anger                     & 0.528 & 1.000 & 0.000 & 1.000 & 0.499 \\
		Disgust                   & 0.025 & 0.000 & 0.000 & 1.000 & 0.156 \\
		Enthusiasm                & 0.253 & 0.000 & 0.000 & 1.000 & 0.435 \\
		Fear                      & 0.075 & 0.000 & 0.000 & 1.000 & 0.263 \\
		Hope                      & 0.072 & 0.000 & 0.000 & 1.000 & 0.258 \\
		Joy                       & 0.272 & 0.000 & 0.000 & 1.000 & 0.445 \\
		Pride                     & 0.109 & 0.000 & 0.000 & 1.000 & 0.312 \\
		Sadness                   & 0.069 & 0.000 & 0.000 & 1.000 & 0.253 \\
        Relatability: Medium      & 0.498 & 0.000 & 0.000 & 1.000 & 0.500 \\
		Relatability: High        & 0.303 & 0.000 & 0.000 & 1.000 & 0.460 \\
        Identity Language         & 0.666 & 1.000 & 0.000 & 1.000 & 0.472 \\
        Outgroup Animosity        & 0.494 & 0.000 & 0.000 & 1.000 & 0.500 \\
		Language Complexity       & 27.234 & 18.095 & 0.400 & 431.882 & 23.982 \\
        Duration            & 1.089 & 0.833 & 0.033 & 59.750 & 1.361 \\
        Pace                      & 2.521 & 2.557 & 0.020 & 23.833 & 0.616 \\ \\
		\midrule
		\multicolumn{6}{l}{\underline{Political Party}}                                                                                                                                                        \\
		SPD                       & 0.256 & 0.000 & 0.000 & 1.000 & 0.436 \\ 
		Die Linke                 & 0.069 & 0.000 & 0.000 & 1.000 & 0.253 \\
		Grüne                     & 0.217 & 0.000 & 0.000 & 1.000 & 0.412 \\
		CDU/CSU                   & 0.163 & 0.000 & 0.000 & 1.000 & 0.369 \\
		FDP                       & 0.072 & 0.000 & 0.000 & 1.000 & 0.258 \\
		AfD                       & 0.224 & 0.000 & 0.000 & 1.000 & 0.417 \\
		\addlinespace
		\multicolumn{6}{l}{\underline{Political Alignment}}                                                                                                                                                  \\
		Far-Left                  & 0.069 & 0.000 & 0.000 & 1.000 & 0.253 \\
		Center-Left               & 0.473 & 0.000 & 0.000 & 1.000 & 0.499 \\
		Center-Right              & 0.235 & 0.000 & 0.000 & 1.000 & 0.424 \\
		Far-Right                 & 0.224 & 0.000 & 0.000 & 1.000 & 0.417 \\
		\bottomrule
		\addlinespace
	\end{tabular}
	}
	\label{tbl:descriptives}
\end{table}

\section{LLM-Based Annotation}
\label{supp:llm_annotation}

To annotate group-oriented messaging (\ie, \textit{Identity Language}, \textit{Outgroup Animosity}, and \textit{Relatability}) in TikTok videos, we implemented an OpenAI Assistant (gpt-4-turbo). The corresponding prompt is printed below.

{\itshape\singlespacing

``You are a professional annotator specializing in political discourse analysis. Your task is to analyze short TikTok captions in German related to the 2025 German federal elections (Bundestagswahl). Each caption must be annotated along three distinct dimensions: identity language, outgroup animosity, and relatability.

Your goal is to produce consistent, structured annotations. Follow the definitions and formatting instructions strictly. Return only a valid JSON object with exactly three top-level keys: 'identity\_language', 'outgroup\_animosity', and 'relatability'.

\smallskip
\noindent

\begin{enumerate}
    \item Identity Language
    \begin{itemize}
        \item  Definition: Identify whether the caption explicitly refers to a political, social, or demographic group commonly associated with German political discourse. These identities may be defined by political affiliation, ethnicity, religion, gender, age cohort, or geographic region.
        \item Examples of identity-relevant groups include (but are not limited to):
        \begin{itemize}
            \item Left-leaning: youth voters (18–24), students, urban residents (e.g., Berlin, Hamburg), individuals with migration backgrounds
            \item Right-leaning: older voters (60+), rural populations, Bavarians, East Germans (e.g., Saxony, Thuringia, Brandenburg), AfD supporters
            \item Religious minorities: Muslims, Jews
            \item Gender groups: women, non-binary people
            \item Ethnic minorities: people of Turkish descent, Eastern European immigrants
        \end{itemize}
        \item Output format:
        \begin{itemize}
            \item If no identity language is detected: "identity\_language": "No"
            \item  If identity language is present:
      "identity\_language": { \\
        "present": "Yes", \\
        "identity\_group": "youth voters",   ()                   // Any specific identity group explicitly referenced \\
        "category": "Generational",                           // One of: Political, Ethnic, Religious, Gender, Generational, Regional \\
        "tone": "Neutral",                                    // One of: Positive, Neutral, Negative \\
        "perspective": "Out-group",                           // One of: In-group, Out-group, Neutral \\
        "emphasizes\_differences\_or\_conflict": "Yes"           // Yes or No \\
      }
        \end{itemize}

    \end{itemize}
   \item Outgroup Animosity
   \begin{itemize}
       \item Definition: Outgroup animosity refers to explicit negative characterizations of political or ideological opponents. This includes emotionally charged or polarizing language that reinforces social identity divides.
        \item Indicators include:
        \begin{itemize}
            \item Direct hostile references to political parties or ideological opponents
            \item Language expressing fear, anger, contempt, or mockery
            \item Claims of moral or existential threat posed by the out-group
        \end{itemize}
        \item Common out-groups may include: AfD, CDU, Die Grünen, leftists, conservatives, elites, immigrants, mainstream media, etc.
        \item Output format: 
        \begin{itemize}
            \item If no outgroup animosity is present: "outgroup\_animosity": "No"
            \item If outgroup animosity is present: \\
      "outgroup\_animosity": { \\
        "present": "Yes", \\
        "identity": "The Greens",                            // Any group that is the target of animosity \\
        "emphasizes\_conflict\_or\_threat": "Yes"               // Yes or No \\
      }
        \end{itemize}
   \end{itemize}
    \item Relatability
    \begin{itemize}
        \item  Definition: Relatability refers to how likely it is that an average German voter would perceive the content as authentic, familiar, and emotionally resonant. It is a measure of connection, identification, and trust.
        \item Key indicators:
        \begin{itemize}
            \item Authenticity: Honest, direct, and credible tone
            \item Shared experience: Use of language, themes, or stories that reflect everyday life or common concerns
            \item Emotional connection: Establishes a sense of personal rapport or identification
        \end{itemize}
        \item  Output format: "relatability": "Medium"  // One of: Low, Medium, High
    \end{itemize}
\end{enumerate}

Return only a valid JSON object using the structure above. Do not include any explanation, comments, or metadata."
}

\section{Validation User Study}
\label{supp:validation}

We conducted a user study on Prolific to validate the automated identification of content features. Twenty native German-speaking participants (9 female; 11 male; mean age of $39.55$ years) were recruited and presented with a randomly selected sample of 100 videos. Among them, ten held a bachelor's degree, nine held a master's degree, and one had a doctoral degree. 

For each video, participants were asked to indicate whether the video contained \textit{Identity Language} (Yes/No), \textit{Outgroup Animosity} (Yes/No), the level of perceived \textit{Relatability} (low, medium, or high), as well as the presence (Yes/No) of eight discrete emotions (\ie \textit{Anger, Disgust, Enthusiasm, Fear, Hope, Joy, Pride, and Sadness}) and the overall \textit{Sentiment} (on a scale from negative to positive). 

\Cref{tbl:user_study} shows the mean participant ratings for each content characteristic, the odds ratio and the associated $p$-values for binary variables (Fisher’s exact test), $p$-values for ordinal variables (Kruskal-Wallis test), and the intraclass correlation coefficient (ICC), which measures interrater reliability \cite{Koo.2016}.
Across all binary content characteristics, mean human ratings were significantly (all $p <0.05$) higher when the characteristic was labeled as present by the LLM (indicated as \textit{Yes}) than if was labeled as absent (\textit{No}), indicating strong alignment between human judgements and LLM annotations. 
For ordinal content characteristics such as \textit{Relatability} and \textit{Sentiment}, human ratings increased monotonically with characteristic levels (all $p<0.01$), such that higher \textit{Relatability} or more positive \textit{Sentiment} corresponded to higher mean ratings, and lower levels corresponded to lower means.
The human raters showed moderate to high inter-rater reliability (ICC ranging from $0.345$ to $0.814$). Overall, these results validate the annotation process and confirm that it captures content features in a way that closely aligns with human assessments.

\begin{table}[H]
    \caption{Mean ratings of raters, odds ratio and corresponding $p$-values for binary variables (Fisher’s exact test), $p$-values for ordinal variables (Kruskal-Wallis test), and interrater reliability (ICC) for each content characteristic. 
    }
    \centering
    \scriptsize
    \begin{tabular}{l l S[table-format=1.3] l l l}
        \toprule
        \textbf{Characteristics} & \textbf{Level} & \textbf{Mean} & \textbf{Odds Ratio} & \textbf{$\bm{p}$-value}& \textbf{ICC} \\
        \midrule
        \multirow{3}{*}{Sentiment}          & Negative   & -0.356  &  \multirow{3}{*}{$-$}       & \multirow{3}{*}{$p < 0.001$}         & \multirow{3}{*}{0.895}           \\
                  & Neutral    & -0.018   &        &         &            \\
                  & Positive   & 0.279    &        &         &            \\
        \midrule
        \multirow{2}{*}{Anger}              & Yes        & 0.405  &  \multirow{2}{*}{6.448}         & \multirow{2}{*}{$p < 0.001$}        & \multirow{2}{*}{0.814}           \\
                      & No         &  0.096   &        &         &            \\
        \midrule
        \multirow{2}{*}{Disgust}            & Yes        & 0.154  &   \multirow{2}{*}{3.356}        & \multirow{2}{*}{$p < 0.05$}        & \multirow{2}{*}{0.551}           \\
                    & No         & 0.051   &         &         &            \\
        \midrule
        \multirow{2}{*}{Enthusiasm}         & Yes        & 0.538  &  \multirow{2}{*}{2.373}        & \multirow{2}{*}{$p < 0.001$}         & \multirow{2}{*}{0.780}           \\
                 & No         & 0.330    &        &         &            \\
        \midrule
        \multirow{2}{*}{Fear}               & Yes        & 0.311  & \multirow{2}{*}{2.927}        & \multirow{2}{*}{$p < 0.001$}         & \multirow{2}{*}{0.560}           \\
                       & No         & 0.134   &         &         &            \\
        \midrule
        \multirow{2}{*}{Hope}                & Yes        & 0.410  &  \multirow{2}{*}{1.854}         & \multirow{2}{*}{$p < 0.01$}         & \multirow{2}{*}{0.641}           \\
                        & No         & 0.270   &         &         &            \\
        \midrule
        \multirow{2}{*}{Joy}               & Yes        & 0.245  &  \multirow{2}{*}{2.335}         & \multirow{2}{*}{$p < 0.001$}         & \multirow{2}{*}{0.781}           \\
                       & No         & 0.121   &         &         &            \\
        \midrule
        \multirow{2}{*}{Pride}              & Yes        & 0.263  &  \multirow{2}{*}{3.147}         & \multirow{2}{*}{$p < 0.001$}         & \multirow{2}{*}{0.542}           \\
                      & No         & 0.103    &        &         &            \\
        \midrule
        \multirow{2}{*}{Sadness}            & Yes        & 0.259  &  \multirow{2}{*}{3.758}         & \multirow{2}{*}{$p < 0.001$}         & \multirow{2}{*}{0.640}           \\
                    & No         & 0.085    &        &         &         \\ 
        \midrule
        \multirow{3}{*}{Relatability}       & Low        & 0.392  &  \multirow{3}{*}{$-$}        & \multirow{3}{*}{$p < 0.01$}        & \multirow{3}{*}{0.680}           \\
                & Medium     & 0.458    &        &         &            \\
                & High       & 0.518     &       &         &            \\
        \midrule
        \multirow{2}{*}{Identity Language}  & Yes & 0.462 & \multirow{2}{*}{2.573}    & \multirow{2}{*}{$p < 0.001$} & \multirow{2}{*}{0.345} \\
          & No  & 0.249 &      &                         &                       \\
        \midrule
        \multirow{2}{*}{Outgroup Animosity}  & Yes        & 0.456   &  \multirow{2}{*}{7.746}       & \multirow{2}{*}{$p < 0.001$}        & \multirow{2}{*}{0.715}           \\
         & No         & 0.097     &       &         &            \\
        \bottomrule
    \end{tabular}
    \label{tbl:user_study}
\end{table}

\section{Estimation Results}
\label{supp:estimation_app}

\begin{table}[H]
	\caption{Negative binomial regression with the number of (i) likes, (ii) comments, (iii) shares, and (iv) views as dependent variables. For each outcome variable, we estimate two models: one that includes political parties and the other that includes political alignment. Account-specific random effects and monthly fixed effects are included.}
	\label{tbl:regression_dv}
	{
		\tiny
		\begin{tabularx}{\textwidth}{@{\hspace{\tabcolsep}\extracolsep{\fill}}l *{8}{S}}
			\toprule
			                      & \multicolumn{2}{c}{\textbf{DV: Like Count}} & \multicolumn{2}{c}{\textbf{DV: Comment Count}} & \multicolumn{2}{c}{\textbf{DV: Share Count}} & \multicolumn{2}{c}{\textbf{DV: View Count}}                                                                     \\
			                      & {\textbf{Model (1)}}                        & {\textbf{Model (2)}}                        & {\textbf{Model (1)}}                           & {\textbf{Model (2)}} & {\textbf{Model (1)}} & {\textbf{Model (2)}} & {\textbf{Model (1)}} & {\textbf{Model (2)}} \\
			\midrule
			\underline{Content Characteristics}                                                                                                                                                                                                     \\
			Sentiment: Positive      & -0.116^{***} & -0.116^{***} & -0.062^{*}   & -0.062^{*}   & -0.165^{***} & -0.165^{***} & -0.110^{***} & -0.111^{***} \\
                                     & (0.022)      & (0.022)      & (0.027)      & (0.027)      & (0.037)      & (0.037)      & (0.023)      & (0.023)      \\
			Sentiment: Negative      & 0.110^{***}  & 0.110^{***}  & 0.116^{***}  & 0.116^{***}  & 0.113^{**}   & 0.114^{**}   & 0.080^{***}  & 0.080^{***}  \\
                                     & (0.024)      & (0.024)      & (0.029)      & (0.029)      & (0.038)      & (0.038)      & (0.024)      & (0.024)      \\
			Anger                    & 0.158^{***}  & 0.158^{***}  & 0.048        & 0.048        & 0.140^{***}  & 0.140^{***}  & 0.084^{***}  & 0.084^{***}  \\
                                     & (0.021)      & (0.021)      & (0.025)      & (0.025)      & (0.034)      & (0.034)      & (0.021)      & (0.021)      \\
			Disgust                  & 0.421^{***}  & 0.421^{***}  & 0.519^{***}  & 0.519^{***}  & 0.685^{***}  & 0.685^{***}  & 0.404^{***}  & 0.404^{***}  \\
                                     & (0.055)      & (0.055)      & (0.068)      & (0.068)      & (0.085)      & (0.085)      & (0.056)      & (0.056)      \\
			Enthusiasm               & -0.146^{***} & -0.145^{***} & 0.188^{***}  & 0.189^{***}  & -0.259^{***} & -0.259^{***} & -0.178^{***} & -0.178^{***} \\
                                     & (0.021)      & (0.021)      & (0.025)      & (0.025)      & (0.033)      & (0.033)      & (0.021)      & (0.021)      \\
			Fear                     & -0.069^{*}   & -0.068^{*}   & -0.021       & -0.020       & -0.067       & -0.067       & -0.027       & -0.027       \\
                                     & (0.034)      & (0.034)      & (0.040)      & (0.040)      & (0.053)      & (0.053)      & (0.034)      & (0.034)      \\
			Hope                     & -0.061       & -0.061       & -0.022       & -0.022       & -0.047       & -0.047       & -0.063       & -0.063       \\
                                     & (0.034)      & (0.034)      & (0.040)      & (0.040)      & (0.054)      & (0.054)      & (0.034)      & (0.034)      \\
			Joy                      & -0.009       & -0.010       & -0.207^{***} & -0.208^{***} & -0.038       & -0.038       & -0.038       & -0.038       \\
                                     & (0.021)      & (0.021)      & (0.025)      & (0.025)      & (0.033)      & (0.033)      & (0.021)      & (0.021)      \\
			Pride                    & 0.049        & 0.049        & 0.040        & 0.040        & 0.040        & 0.040        & 0.031        & 0.031        \\
                                     & (0.028)      & (0.028)      & (0.034)      & (0.034)      & (0.046)      & (0.046)      & (0.029)      & (0.029)      \\
			Sadness                  & 0.120^{***}  & 0.120^{***}  & 0.049        & 0.051        & 0.043        & 0.044        & 0.117^{**}   & 0.118^{**}   \\
                                     & (0.036)      & (0.036)      & (0.043)      & (0.043)      & (0.057)      & (0.057)      & (0.036)      & (0.036)      \\
			Relatability: Medium     & -0.365^{***} & -0.364^{***} & -0.222^{***} & -0.222^{***} & -0.632^{***} & -0.632^{***} & -0.302^{***} & -0.302^{***} \\
                                     & (0.025)      & (0.025)      & (0.030)      & (0.030)      & (0.039)      & (0.039)      & (0.025)      & (0.025)      \\
			Relatability: High       & -0.501^{***} & -0.501^{***} & -0.324^{***} & -0.326^{***} & -0.759^{***} & -0.760^{***} & -0.455^{***} & -0.455^{***} \\
                                     & (0.030)      & (0.030)      & (0.036)      & (0.036)      & (0.047)      & (0.047)      & (0.030)      & (0.030)      \\
			Identity Language        & -0.001       & -0.001       & -0.003       & -0.003       & -0.020       & -0.020       & -0.048^{*}   & -0.048^{*}   \\
                                     & (0.020)      & (0.020)      & (0.024)      & (0.024)      & (0.032)      & (0.032)      & (0.020)      & (0.020)      \\
			Outgroup Animosity       & 0.401^{***}  & 0.401^{***}  & 0.583^{***}  & 0.583^{***}  & 0.448^{***}  & 0.448^{***}  & 0.315^{***}  & 0.316^{***}  \\
                                     & (0.022)      & (0.022)      & (0.026)      & (0.026)      & (0.036)      & (0.036)      & (0.022)      & (0.022)      \\
            \addlinespace
			\underline{Structural Characteristics}                                                                                                                                                                                                       \\
            Language Complexity      & 0.110^{***}  & 0.111^{***}  & 0.162^{***}  & 0.162^{***}  & 0.106^{***}  & 0.106^{***}  & 0.123^{***}  & 0.123^{***}  \\
                                     & (0.011)      & (0.011)      & (0.013)      & (0.013)      & (0.017)      & (0.017)      & (0.011)      & (0.011)      \\
			Duration           & -0.032^{**}  & -0.032^{**}  & -0.078^{***} & -0.079^{***} & 0.030        & 0.030        & -0.031^{**}  & -0.031^{**}  \\
                                     & (0.011)      & (0.011)      & (0.011)      & (0.011)      & (0.020)      & (0.020)      & (0.011)      & (0.011)      \\
            Pace                     & 0.008        & 0.008        & 0.038^{**}   & 0.039^{**}   & -0.099^{***} & -0.099^{***} & 0.004        & 0.004        \\
                                     & (0.010)      & (0.010)      & (0.012)      & (0.012)      & (0.015)      & (0.015)      & (0.010)      & (0.010)      \\
			\addlinespace
			\underline{Political Parties}                                                                                                                                                                                                           \\
			Die Linke                & 1.551^{***}  &              & 1.004^{***}  &              & 1.459^{***}  &              & 0.809^{***}  &              \\
                                     & (0.236)      &              & (0.275)      &              & (0.310)      &              & (0.226)      &              \\
			Grüne                    & 0.291^{*}    &              & 0.677^{***}  &              & 0.159        &              & 0.153        &              \\
                                     & (0.147)      &              & (0.172)      &              & (0.195)      &              & (0.141)      &              \\
			CDU/CSU                  & -0.379^{*}   &              & -0.409^{*}   &              & -0.154       &              & -0.123       &              \\
                                     & (0.163)      &              & (0.192)      &              & (0.216)      &              & (0.156)      &              \\
			FDP                      & -0.109       &              & -0.083       &              & 0.159        &              & -0.056       &              \\
                                     & (0.212)      &              & (0.248)      &              & (0.281)      &              & (0.203)      &              \\
			AfD                      & 1.401^{***}  &              & 1.074^{***}  &              & 1.761^{***}  &              & 0.825^{***}  &              \\
                                     & (0.168)      &              & (0.196)      &              & (0.222)      &              & (0.161)      &              \\
			\addlinespace
			\underline{Political Alignment}                                                                                                                                                                                                         \\
			Far-Left                 &              & 1.421^{***}  &              & 0.700^{**}   &              & 1.387^{***}  &              & 0.740^{***}  \\
                                     &              & (0.227)      &              & (0.266)      &              & (0.297)      &              & (0.217)      \\
			Center-Right             &              & -0.422^{**}  &              & -0.609^{***} &              & -0.125       &              & -0.170       \\
                                     &              & (0.130)      &              & (0.154)      &              & (0.171)      &              & (0.124)      \\
			Far-Right                &              & 1.270^{***}  &              & 0.768^{***}  &              & 1.689^{***}  &              & 0.756^{***}  \\
                                     &              & (0.155)      &              & (0.183)      &              & (0.204)      &              & (0.148)      \\
			\addlinespace
		\end{tabularx}
	}
\end{table}

\newpage

\begin{table}[H]
	\label{tbl:regression_dv2}
	{
		\tiny
		\begin{tabularx}{\textwidth}{@{\hspace{\tabcolsep}\extracolsep{\fill}}l *{8}{S}}
			\underline{Additional Controls}                                                                                                                                                                                                         \\
			User Random Effects   & {Included}                                  & {Included}                                  & {Included}                                     & {Included}           & {Included}           & {Included}    & {Included} & {Included}       \\
			Monthly Fixed Effects & {Included}                                  & {Included}                                  & {Included}                                     & {Included}           & {Included}           & {Included}      &  {Included} & {Included}   \\
			\midrule
			AIC                   & {\num{353747}}                              & {\num{353748}}                              & {\num{238584}}                                 & {\num{238597}}       & {\num{189928}}       & {\num{189926}} &   {\num{499339}}       & {\num{499336}}     \\
			Observations          & {\num{25292}}                               & {\num{25292}}                               & {\num{25292}}                                  & {\num{25292}}        & {\num{25292}}        & {\num{25292}} &    {\num{25292}}        & {\num{25292}}   \\
			\bottomrule
            \multicolumn{9}{l}{\tiny{Significance levels: $^{***}p<0.001$; $^{**}p<0.01$; $^{*}p<0.05$}} \\
		\end{tabularx}
	}
\end{table}

\newpage
\section{Exploratory Analyses and Robustness Checks}
\label{supp:robustness}

\subsection{Variance Inflation Factors}

We calculated variance inflation factors for all explanatory variables in our regression models, separately for each outcome and political affiliation variable (\Cref{tbl:vif}). The VIFs are substantially below the critical threshold of four, which indicates that multicollinearity is not an issue in our analysis.

\begin{table}[H]
	\caption{Variance Inflation Factors for Regression Models.}
	\label{tbl:vif}
	{
		\tiny
		\begin{tabularx}{\textwidth}{@{\hspace{\tabcolsep}\extracolsep{\fill}}l *{8}{S}}
			\toprule
			                      & \multicolumn{2}{c}{\textbf{DV: Like Count}} & \multicolumn{2}{c}{\textbf{DV: Comment Count}} & \multicolumn{2}{c}{\textbf{DV: Share Count}} & \multicolumn{2}{c}{\textbf{DV: View Count}}                                                                     \\
			                      & {\textbf{Model (1)}}                        & {\textbf{Model (2)}}                        & {\textbf{Model (1)}}                           & {\textbf{Model (2)}} & {\textbf{Model (1)}} & {\textbf{Model (2)}} & {\textbf{Model (1)}} & {\textbf{Model (2)}} \\
			\midrule
			Sentiment                & 1.22         & 1.22         & 1.23         & 1.23         & 1.25         & 1.25         & 1.22         & 1.22         \\
			Anger                    & 1.32         & 1.32         & 1.31         & 1.31         & 1.35         & 1.35         & 1.31         & 1.31         \\
			Disgust                  & 1.05         & 1.05         & 1.05         & 1.05         & 1.05         & 1.05         & 1.05         & 1.05         \\
			Enthusiasm               & 1.10         & 1.10         & 1.09         & 1.09         & 1.09         & 1.10         & 1.10         & 1.10         \\
			Fear                     & 1.07         & 1.07         & 1.06         & 1.06         & 1.09         & 1.09         & 1.07         & 1.07         \\
			Hope                     & 1.06         & 1.06         & 1.06         & 1.06         & 1.05         & 1.05         & 1.05         & 1.05         \\
			Joy                      & 1.13         & 1.13         & 1.13         & 1.13         & 1.13         & 1.13         & 1.12         & 1.12         \\
			Pride                    & 1.08         & 1.08         & 1.09         & 1.09         & 1.09         & 1.09         & 1.08         & 1.08         \\
			Sadness                  & 1.08         & 1.08         & 1.07         & 1.07         & 1.10         & 1.10         & 1.08         & 1.08         \\
			Relatability             & 1.15         & 1.15         & 1.16         & 1.16         & 1.15         & 1.16         & 1.14         & 1.14         \\
			Identity Language        & 1.14         & 1.14         & 1.14         & 1.14         & 1.17         & 1.17         & 1.15         & 1.15         \\
			Outgroup Animosity       & 1.33         & 1.33         & 1.32         & 1.32         & 1.35         & 1.35         & 1.33         & 1.33         \\
            Language Complexity      & 1.17         & 1.17         & 1.16         & 1.16         & 1.19         & 1.20         & 1.17         & 1.17         \\
			Duration           & 1.26         & 1.26         & 1.22         & 1.22         & 1.30         & 1.30         & 1.25         & 1.25         \\
            Pace                     & 1.07         & 1.07         & 1.08         & 1.08         & 1.08         & 1.08         & 1.07         & 1.07         \\
			Alignment                & 1.01         &              & 1.01         &              & 1.01         &              & 1.01         &              \\
			Party                    &              & 1.01         &              & 1.01         &              & 1.02         &              & 1.01         \\
			\addlinespace
			\bottomrule
		\end{tabularx}
	}
\end{table}

\subsection{Comparison to an Implausible Model}

To validate our findings, we repeated our analysis using an implausible XYZ model  \cite{Feuerriegel.2025}. This was done to address the concern that observational studies can sometimes produce evidence for absurd models \cite{Burton.2021,Solovev.2023}. For each video, we counted the number of X's, Y's and Z's (\ie, an absurd factor) and used this as an explanatory variable to test its ability to predict the amount of likes, comments, shares, and views a video received. 

The regression results (see \Cref{tbl:regression_xyz}) showed that, across all four outcome variables, the XYZ model had a higher AIC value compared to our original model, indicating a lower model adequacy. Furthermore, the effect sizes of the \textit{XYZ} variables were negligible, with coefficient estimates close to zero. Specifically, the coefficients for the \textit{XYZ} variables were $-0.005$ ($p<0.01$) for the number of likes, $-0.004$ for comments ($p>0.05$), $-0.010$ ($p<0.001$) for shares, and $-0.008$ ($p<0.001$) for views, and. For comparison, the coefficients of the content characteristic \textit{Outgroup Animosity} were $0.401$ ($p<0.001$) for likes, $0.583$ ($p<0.001$) for comments, $0.448$ ($p<0.001$) for shares, and $0.080$ ($p<0.315$) for views ($p<0.001$), \ie, drastically larger. 
Overall, these results provide strong confirmatory evidence that the content characteristics used in our model were meaningful predictors of user engagement.

\begin{table}[H]
	\caption{Comparison to an implausible XYZ model \cite{Feuerriegel.2025}. The variable \textit{XYZ} measures the number of X’s, Y’s and Z’s in the videos (\ie, an absurd factor).}
	\label{tbl:regression_xyz}
	{
		\tiny
		\begin{tabularx}{\textwidth}{@{\hspace{\tabcolsep}\extracolsep{\fill}}l *{8}{S}}
			\toprule
			                      & \multicolumn{2}{c}{\textbf{DV: Like Count}} & \multicolumn{2}{c}{\textbf{DV: Comment Count}} & \multicolumn{2}{c}{\textbf{DV: Share Count}} & \multicolumn{2}{c}{\textbf{DV: View Count}}                                                                     \\
			                      & {\textbf{Model (1)}}                        & {\textbf{Model (2)}}                        & {\textbf{Model (1)}}                           & {\textbf{Model (2)}} & {\textbf{Model (1)}} & {\textbf{Model (2)}} & {\textbf{Model (1)}} & {\textbf{Model (2)}} \\
			\midrule
			\underline{Control Variables}                                                                                                                                                                                                     \\
            XYZ                      & -0.005^{**} & -0.005^{**} & -0.004      & -0.004       & -0.010^{***} & -0.010^{***} & -0.008^{***} & -0.008^{***} \\
                                     & (0.002)     & (0.002)     & (0.002)     & (0.002)      & (0.003)      & (0.003)      & (0.002)      & (0.002)      \\ 
            Language Complexity      & 0.127^{***} & 0.127^{***} & 0.198^{***} & 0.199^{***}  & 0.110^{***}  & 0.110^{***}  & 0.131^{***}  & 0.131^{***}  \\
                                     & (0.011)     & (0.011)     & (0.013)     & (0.013)      & (0.017)      & (0.017)      & (0.011)      & (0.011)      \\
			Duration           & 0.087^{***} & 0.086^{***} & 0.023       & 0.022        & 0.224^{***}  & 0.224^{***}  & 0.093^{***}  & 0.093^{***}  \\
                                     & (0.026)     & (0.026)     & (0.031)     & (0.031)      & (0.041)      & (0.041)      & (0.025)      & (0.025)      \\
            Pace                     & 0.016       & 0.016       & 0.051^{***} & 0.051^{***}  & -0.097^{***} & -0.097^{***} & 0.010        & 0.010        \\
                                     & (0.010)     & (0.010)     & (0.013)     & (0.013)      & (0.016)      & (0.016)      & (0.010)      & (0.010)      \\
			\addlinespace
			\underline{Political Parties}                                                                                                                                                                                                           \\
			Die Linke                & 1.646^{***} &             & 1.151^{***} &              & 1.581^{***}  &              & 0.900^{***}  &              \\
                                     & (0.247)     &             & (0.284)     &              & (0.324)      &              & (0.234)      &              \\
			Grüne                    & 0.360^{*}   &             & 0.744^{***} &              & 0.257        &              & 0.212        &              \\
                                     & (0.154)     &             & (0.179)     &              & (0.204)      &              & (0.146)      &              \\
			CDU/CSU                  & -0.338^{*}  &             & -0.370      &              & -0.062       &              & -0.083       &              \\
                                     & (0.171)     &             & (0.198)     &              & (0.226)      &              & (0.162)      &              \\
			FDP                      & -0.024      &             & -0.005      &              & 0.307        &              & 0.037        &              \\
                                     & (0.222)     &             & (0.257)     &              & (0.294)      &              & (0.211)      &              \\
			AfD                      & 1.792^{***} &             & 1.414^{***} &              & 2.324^{***}  &              & 1.159^{***}  &              \\
                                     & (0.176)     &             & (0.203)     &              & (0.232)      &              & (0.167)      &              \\
			\addlinespace
			\underline{Political Alignment}                                                                                                                                                                                                         \\
			Far-Left                 &             & 1.484^{***} &             & 0.816^{**}   &              & 1.465^{***}  &              & 0.805^{***}  \\
                                     &             & (0.239)     &             & (0.276)      &              & (0.312)      &              & (0.225)      \\
			Center-Right             &             & -0.398^{**} &             & -0.587^{***} &              & -0.059       &              & -0.139       \\
                                     &             & (0.136)     &             & (0.159)      &              & (0.179)      &              & (0.129)      \\
			Far-Right                &             & 1.630^{***} &             & 1.078^{***}  &              & 2.207^{***}  &              & 1.064^{***}  \\
                                     &             & (0.163)     &             & (0.189)      &              & (0.213)      &              & (0.154)      \\
			\addlinespace
			\underline{Additional Controls}                                                                                                                                                                                                         \\
			User Random Effects   & {Included}                                  & {Included}                                  & {Included}                                     & {Included}           & {Included}           & {Included}    & {Included} & {Included}       \\
			Monthly Fixed Effects & {Included}                                  & {Included}                                  & {Included}                                     & {Included}           & {Included}           & {Included}      &  {Included} & {Included}   \\
			\midrule
			AIC                   & {\num{355235}}                              & {\num{355238}}                              & {\num{239745}}                                 & {\num{239759}}       & {\num{190920}}       & {\num{190919}} &   {\num{500352}}       & {\num{500351}}     \\
			Observations          & {\num{25292}}                               & {\num{25292}}                               & {\num{25292}}                                  & {\num{25292}}        & {\num{25292}}        & {\num{25292}} &    {\num{25292}}        & {\num{25292}}   \\
			\bottomrule
            \multicolumn{9}{l}{\tiny{Significance levels: $^{***}p<0.001$; $^{**}p<0.01$; $^{*}p<0.05$}} \\
		\end{tabularx}
	}
\end{table}
\newpage

\subsection{Analysis With Continuous Sentiment and Emotion Scores}

In the main model, \textit{Sentiment} was coded as a categorical variable (\ie, negative, neutral [reference], positive), and discrete emotions (\ie, \textit{Anger}, \textit{Disgust}, \textit{Enthusiasm}, \textit{Fear}, \textit{Hope}, \textit{Joy}, \textit{Pride}, and \textit{Sadness}) were represented as binary indicators denoting their presence in a video. As a robustness check, we instead operationalized \textit{Sentiment} as a continuous variable, defined as the mean sentiment score of a video. For discrete emotions, we calculated continuous intensity measures by computing the proportion of sentences in a video expressing each emotion relative to the total number of sentences. All sentiment and emotion measures were $z$-standardized for comparability. The regression results are reported in \Cref{tbl:regression_continuous}.

The continuous specification yielded results closely aligned with the main model: most coefficients retained their sign, magnitude, and statistical significance. Continuous sentiment was negatively associated with all engagement metrics (all $p<0.001$), with a one standard deviation increase corresponding to a \SI{6.4}{\percent} decrease in likes (coef: $-0.066$), \SI{3.9}{\percent} decrease in comments (coef: $-0.040$), \SI{6.9}{\percent} decrease in shares (coef: $-0.072$), and \SI{5.2}{\percent} decrease in views (coef: $-0.054$). This mirrors the original finding that positive sentiment reduced, and negative sentiment increased, engagement. Results for discrete emotions were also largely consistent, with only minor changes: \textit{Fear} lost its small but statistically significant negative effect, \textit{Hope} now shows small but statistically significant negative associations with likes and views, and \textit{Sadness} additionally became significant for comments. All other content features, including \textit{Relatability}, \textit{Identity Language}, \textit{Outgroup Animosity}, party affiliation, and ideological alignment, remained robust. Overall, these results demonstrate that our conclusions are robust to alternative operationalizations of sentiment and emotion.

\begin{table}[H]
	\caption{Negative binomial regression with the number of (i) likes, (ii) comments, (iii) shares, and (iv) views as dependent variables.  For each outcome variable, we estimate two models: one that includes political parties and the other that includes political alignment. Account-specific random effects and monthly fixed effects are included. }
	\label{tbl:regression_continuous}
	{
		\tiny
		\begin{tabularx}{\textwidth}{@{\hspace{\tabcolsep}\extracolsep{\fill}}l *{8}{S}}
			\toprule
			                      & \multicolumn{2}{c}{\textbf{DV: Like Count}} & \multicolumn{2}{c}{\textbf{DV: Comment Count}} & \multicolumn{2}{c}{\textbf{DV: Share Count}} & \multicolumn{2}{c}{\textbf{DV: View Count}}                                                                     \\
			                      & {\textbf{Model (1)}}                        & {\textbf{Model (2)}}                        & {\textbf{Model (1)}}                           & {\textbf{Model (2)}} & {\textbf{Model (1)}} & {\textbf{Model (2)}} & {\textbf{Model (1)}} & {\textbf{Model (2)}} \\
			\midrule
			\underline{Content Characteristics}                                                                                                                                                                                                     \\
			Sentiment                & -0.066^{***} & -0.066^{***} & -0.040^{***} & -0.040^{***} & -0.072^{***} & -0.072^{***} & -0.054^{***} & -0.054^{***} \\
                                     & (0.010)      & (0.010)      & (0.012)      & (0.012)      & (0.016)      & (0.016)      & (0.010)      & (0.010)      \\
			Anger                    & 0.039^{***}  & 0.039^{***}  & 0.040^{**}   & 0.040^{**}   & 0.002        & 0.002        & 0.006        & 0.006        \\
                                     & (0.011)      & (0.011)      & (0.012)      & (0.012)      & (0.017)      & (0.017)      & (0.010)      & (0.010)      \\
			Disgust                  & 0.062^{***}  & 0.062^{***}  & 0.070^{***}  & 0.070^{***}  & 0.096^{***}  & 0.096^{***}  & 0.067^{***}  & 0.067^{***}  \\
                                     & (0.010)      & (0.010)      & (0.012)      & (0.012)      & (0.016)      & (0.016)      & (0.010)      & (0.010)      \\
			Enthusiasm               & -0.055^{***} & -0.054^{***} & 0.104^{***}  & 0.104^{***}  & -0.101^{***} & -0.101^{***} & -0.059^{***} & -0.059^{***} \\
                                     & (0.008)      & (0.008)      & (0.011)      & (0.011)      & (0.013)      & (0.013)      & (0.008)      & (0.008)      \\
			Fear                     & -0.008       & -0.008       & 0.014        & 0.014        & -0.016       & -0.016       & 0.006        & 0.006        \\
                                     & (0.008)      & (0.008)      & (0.010)      & (0.010)      & (0.013)      & (0.013)      & (0.008)      & (0.008)      \\
			Hope                     & -0.035^{***} & -0.035^{***} & -0.014       & -0.014       & -0.025       & -0.025       & -0.025^{**}  & -0.025^{**}  \\
                                     & (0.008)      & (0.008)      & (0.010)      & (0.010)      & (0.014)      & (0.014)      & (0.008)      & (0.008)      \\
			Joy                      & -0.003       & -0.003       & -0.042^{***} & -0.042^{***} & -0.032^{*}   & -0.032^{*}   & -0.007       & -0.007       \\
                                     & (0.009)      & (0.009)      & (0.011)      & (0.011)      & (0.015)      & (0.015)      & (0.009)      & (0.009)      \\
			Pride                    & -0.006       & -0.006       & 0.007        & 0.007        & -0.028       & -0.028       & -0.014       & -0.014       \\
                                     & (0.009)      & (0.009)      & (0.011)      & (0.011)      & (0.015)      & (0.015)      & (0.009)      & (0.009)      \\
			Sadness                  & 0.030^{***}  & 0.031^{***}  & 0.030^{**}   & 0.030^{**}   & 0.004        & 0.004        & 0.031^{***}  & 0.031^{***}  \\
                                     & (0.009)      & (0.009)      & (0.011)      & (0.011)      & (0.015)      & (0.015)      & (0.009)      & (0.009)      \\
			Relatability: Medium     & -0.363^{***} & -0.363^{***} & -0.221^{***} & -0.221^{***} & -0.641^{***} & -0.641^{***} & -0.305^{***} & -0.305^{***} \\
                                     & (0.025)      & (0.025)      & (0.030)      & (0.030)      & (0.039)      & (0.039)      & (0.025)      & (0.025)      \\
			Relatability: High       & -0.497^{***} & -0.498^{***} & -0.322^{***} & -0.324^{***} & -0.769^{***} & -0.770^{***} & -0.463^{***} & -0.464^{***} \\
                                     & (0.030)      & (0.030)      & (0.036)      & (0.036)      & (0.047)      & (0.047)      & (0.030)      & (0.030)      \\
			Identity Language        & -0.010       & -0.010       & -0.005       & -0.005       & -0.038       & -0.038       & -0.058^{**}  & -0.058^{**}  \\
                                     & (0.020)      & (0.020)      & (0.024)      & (0.024)      & (0.032)      & (0.032)      & (0.020)      & (0.020)      \\
			Outgroup Animosity       & 0.425^{***}  & 0.425^{***}  & 0.586^{***}  & 0.586^{***}  & 0.487^{***}  & 0.487^{***}  & 0.334^{***}  & 0.334^{***}  \\
                                     & (0.022)      & (0.022)      & (0.026)      & (0.026)      & (0.035)      & (0.035)      & (0.022)      & (0.022)      \\
            \addlinespace
			\underline{Structural Characteristics}                                                                                                                                                                                                       \\
            Language Complexity      & 0.107^{***}  & 0.108^{***}  & 0.157^{***}  & 0.157^{***}  & 0.106^{***}  & 0.107^{***}  & 0.124^{***}  & 0.124^{***}  \\
                                     & (0.011)      & (0.011)      & (0.013)      & (0.013)      & (0.017)      & (0.017)      & (0.011)      & (0.011)      \\
			Duration           & -0.015       & -0.016       & -0.063^{***} & -0.063^{***} & 0.035^{*}    & 0.035^{*}    & -0.027^{**}  & -0.027^{**}  \\
                                     & (0.010)      & (0.010)      & (0.011)      & (0.011)      & (0.018)      & (0.018)      & (0.010)      & (0.010)      \\
            Pace                     & 0.010        & 0.010        & 0.040^{***}  & 0.040^{***}  & -0.102^{***} & -0.102^{***} & 0.004        & 0.004        \\
                                     & (0.010)      & (0.010)      & (0.012)      & (0.012)      & (0.015)      & (0.015)      & (0.010)      & (0.010)      \\
			\addlinespace
			\underline{Political Parties}                                                                                                                                                                                                           \\
			Die Linke                & 1.559^{***}  &              & 1.006^{***}  &              & 1.470^{***}  &              & 0.817^{***}  &              \\
                                     & (0.237)      &              & (0.275)      &              & (0.311)      &              & (0.226)      &              \\
			Grüne                    & 0.299^{*}    &              & 0.669^{***}  &              & 0.170        &              & 0.158        &              \\
                                     & (0.148)      &              & (0.173)      &              & (0.195)      &              & (0.141)      &              \\
			CDU/CSU                  & -0.373^{*}   &              & -0.416^{*}   &              & -0.139       &              & -0.118       &              \\
                                     & (0.164)      &              & (0.192)      &              & (0.216)      &              & (0.156)      &              \\
			FDP                      & -0.102       &              & -0.087       &              & 0.179        &              & -0.049       &              \\
                                     & (0.213)      &              & (0.249)      &              & (0.281)      &              & (0.204)      &              \\
			AfD                      & 1.418^{***}  &              & 1.072^{***}  &              & 1.795^{***}  &              & 0.842^{***}  &              \\
                                     & (0.169)      &              & (0.197)      &              & (0.223)      &              & (0.162)      &              \\
			\addlinespace
			\underline{Political Alignment}                                                                                                                                                                                                         \\
			Far-Left                 &              & 1.425^{***}  &              & 0.704^{**}   &              & 1.392^{***}  &              & 0.746^{***}  \\
                                     &              & (0.228)      &              & (0.267)      &              & (0.298)      &              & (0.217)      \\
			Center-Right             &              & -0.420^{**}  &              & -0.611^{***} &              & -0.113       &              & -0.167       \\
                                     &              & (0.130)      &              & (0.154)      &              & (0.172)      &              & (0.124)      \\
			Far-Right                &              & 1.283^{***}  &              & 0.769^{***}  &              & 1.717^{***}  &              & 0.771^{***}  \\
                                     &              & (0.156)      &              & (0.183)      &              & (0.204)      &              & (0.149)      \\
			\addlinespace
			\underline{Additional Controls}                                                                                                                                                                                                         \\
			User Random Effects   & {Included}                                  & {Included}                                  & {Included}                                     & {Included}           & {Included}           & {Included}    & {Included} & {Included}       \\
			Monthly Fixed Effects & {Included}                                  & {Included}                                  & {Included}                                     & {Included}           & {Included}           & {Included}      &  {Included} & {Included}   \\
			\midrule
			AIC                   & {\num{353823}}                              & {\num{353825}}                              & {\num{238617}}                                 & {\num{238629}}       & {\num{190000}}       & {\num{189998}} &   {\num{499393}}       & {\num{499390}}     \\
			Observations          & {\num{25292}}                               & {\num{25292}}                               & {\num{25292}}                                  & {\num{25292}}        & {\num{25292}}        & {\num{25292}} &    {\num{25292}}        & {\num{25292}}   \\
			\bottomrule
            \multicolumn{9}{l}{\tiny{Significance levels: $^{***}p<0.001$; $^{**}p<0.01$; $^{*}p<0.05$}} \\
		\end{tabularx}
	}
\end{table}

\subsection{Analysis of Engagement Per Views}

We implemented a zero-inflated beta regression using the number of likes, comments, and shares \emph{per view} as the dependent variables \cite{Cheng.2024}. This allows us to distinguish active engagement (\ie, likes, comments, and shares) from passive engagement (\ie, views without likes, comments, or shares). Notably, videos on TikTok can still go viral through passive engagement alone, since the algorithm also promotes content based on factors like watch time, repeated views, and completion rates \cite{Boeker.2022,Pan.2023}; even when active interactions (\ie, likes, comments, and shares) remain low. The regression results are reported in \Cref{tbl:regression_alternative}.

The coefficient estimates show a relatively clear pattern: language and emotions tied to moralization or high arousal tend to predict higher engagement per view, while low-arousal emotions play a much smaller role. Among all content features, \textit{Outgroup Animosity} showed the largest effect, increasing likes per view by \SI{12.1}{\percent} (coef: 0.114, $p<0.001$), comments by \SI{20.6}{\percent} (coef: 0.187, $p<0.001$), and shares by \SI{5.0}{\percent} (coef: 0.049, $p<0.001$). \textit{Identity Language} also significantly increased likes per view by \SI{3.7}{\percent} (coef: 0.037, $p<0.001$) and comments by \SI{2.8}{\percent} (coef: 0.028, $p<0.05$). Furthermore, \textit{Anger} increased likes per view by \SI{6.0}{\percent} (coef: 0.058, $p<0.001$) and shares per view by \SI{4.5}{\percent} (coef: 0.044, $p<0.001$), while \textit{Disgust} was associated with a \SI{9.8}{\percent} increase in shares per view (coef: 0.094, $p<0.01$). Together, these findings support prior research showing that moral-emotional content triggers stronger active engagement and social amplification \cite{Brady.2017, Brady.2021, Dillion.2024,Solovev.2022b}.

By contrast, low-arousal emotions were either negatively associated or not statistically significant. \textit{Sadness} decreased comments per view by \SI{5.1}{\percent} (coef: -0.052, $p<0.01$), \textit{Hope} showed no significant effects on any outcome variable (each $p>0.05$), and \textit{Joy} reduced comments per view by \SI{9.3}{\percent} (coef: -0.098, $p<0.001$). These results align well with theories that emotional intensity, rather than valence alone, drives engagement \cite{Rathje.2025}.

Interestingly, \textit{Relatability} showed a more mixed relationship with engagement. \textit{Medium Relatability} increased comments per view by \SI{3.7}{\percent} (coef: 0.037, $p<0.01$) but decreased shares by \SI{11.3}{\percent} (coef: -0.120, $p<0.001$). \textit{High Relatability} showed an even stronger effect on comments, increasing them by \SI{8.0}{\percent} (coef: 0.077, $p<0.001$), but similarly decreased shares by \SI{7.4}{\percent} (coef: -0.077, $p<0.001$). This suggests that highly relatable content may encourage self-expression, leading viewers to comment rather than share.

Beyond content characteristics, we again found that videos of extreme parties (from both sides of the political spectrum) were significantly more likely to trigger active engagement than their centrist counterparts (each $p<0.01$). 

Altogether, our analysis of engagement per views implies that especially moralized language and high-arousal emotions (\eg, \textit{Outgroup Animosity}, \textit{Identity Language}, \textit{Anger}, and \textit{Disgust}) were the strongest predictors of active engagement on TikTok during the 2025 German federal election. In contrast, low-arousal emotions (\eg, \textit{Hope} and \textit{Sadness}) were either negatively associated or not statistically significant. 

\begin{table}[H]
	\caption{Zero-inflated beta regression with the number of (i) likes, (ii) comments and (iii) shares per view as dependent variables.  For each outcome variable, we estimate two models: one that includes political parties and the other that includes political alignment. Account-specific random effects and monthly fixed effects are included. }
	\label{tbl:regression_alternative}
	{
		\tiny
		\begin{tabularx}{\textwidth}{@{\hspace{\tabcolsep}\extracolsep{\fill}}l *{6}{S}}
			\toprule
			                      & \multicolumn{2}{c}{\textbf{DV: Like Count/ View Count}} & \multicolumn{2}{c}{\textbf{DV: Comment Count/ View Count}} & \multicolumn{2}{c}{\textbf{DV: Share Count/ View Count}}                                                                    \\
			                      & {\textbf{Model (1)}}                        & {\textbf{Model (2)}}                        & {\textbf{Model (1)}}                           & {\textbf{Model (2)}} & {\textbf{Model (1)}} & {\textbf{Model (2)}} \\
			\midrule
			\underline{Content Characteristics}                                                                                                                                                                                                     \\
			Sentiment: Positive      & -0.008       & -0.008       & -0.002       & -0.002       & -0.032^{*}   & -0.032^{*}   \\
                                     & (0.008)      & (0.008)      & (0.012)      & (0.012)      & (0.015)      & (0.015)      \\
			Sentiment: Negative      & 0.013        & 0.013        & -0.004       & -0.004       & -0.021       & -0.021       \\
                                     & (0.008)      & (0.008)      & (0.013)      & (0.013)      & (0.014)      & (0.014)      \\
			Anger                    & 0.058^{***}  & 0.058^{***}  & 0.002        & 0.003        & 0.044^{***}  & 0.043^{**}   \\
                                     & (0.008)      & (0.008)      & (0.011)      & (0.011)      & (0.013)      & (0.013)      \\
			Disgust                  & 0.005        & 0.005        & 0.024        & 0.025        & 0.094^{**}   & 0.094^{**}   \\
                                     & (0.019)      & (0.019)      & (0.030)      & (0.030)      & (0.030)      & (0.030)      \\
			Enthusiasm               & 0.046^{***}  & 0.046^{***}  & 0.183^{***}  & 0.183^{***}  & 0.014        & 0.014        \\
                                     & (0.007)      & (0.007)      & (0.011)      & (0.011)      & (0.013)      & (0.013)      \\
			Fear                     & 0.008        & 0.008        & 0.038^{*}    & 0.039^{*}    & -0.017       & -0.017       \\
                                     & (0.012)      & (0.012)      & (0.017)      & (0.017)      & (0.020)      & (0.020)      \\
			Hope                     & -0.001       & -0.001       & 0.025        & 0.026        & -0.028       & -0.028       \\
                                     & (0.012)      & (0.012)      & (0.018)      & (0.018)      & (0.022)      & (0.022)      \\
			Joy                      & 0.005        & 0.005        & -0.098^{***} & -0.098^{***} & 0.006        & 0.006        \\
                                     & (0.008)      & (0.008)      & (0.012)      & (0.012)      & (0.013)      & (0.013)      \\
			Pride                    & 0.018        & 0.018        & 0.031^{*}    & 0.031^{*}    & -0.025       & -0.025       \\
                                     & (0.011)      & (0.011)      & (0.015)      & (0.015)      & (0.019)      & (0.019)      \\
			Sadness                  & 0.006        & 0.006        & -0.052^{**}  & -0.051^{**}  & -0.021       & -0.022       \\
                                     & (0.012)      & (0.012)      & (0.019)      & (0.019)      & (0.021)      & (0.021)      \\
			Relatability: Medium     & -0.007       & -0.007       & 0.037^{**}   & 0.037^{**}   & -0.120^{***} & -0.120^{***} \\
                                     & (0.009)      & (0.009)      & (0.014)      & (0.014)      & (0.015)      & (0.015)      \\
			Relatability: High       & 0.010        & 0.009        & 0.078^{***}  & 0.077^{***}  & -0.077^{***} & -0.076^{***} \\
                                     & (0.011)      & (0.011)      & (0.016)      & (0.016)      & (0.018)      & (0.018)      \\
			Identity Language        & 0.037^{***}  & 0.037^{***}  & 0.028^{*}    & 0.028^{*}    & 0.010        & 0.010        \\
                                     & (0.007)      & (0.007)      & (0.011)      & (0.011)      & (0.013)      & (0.013)      \\
			Outgroup Animosity       & 0.114^{***}  & 0.114^{***}  & 0.187^{***}  & 0.187^{***}  & 0.049^{***}  & 0.049^{***}  \\
                                     & (0.008)      & (0.008)      & (0.012)      & (0.012)      & (0.014)      & (0.014)      \\
            \addlinespace
			\underline{Structural Characteristics}                                                                                                                                                                                                       \\
            Language Complexity      & -0.019^{***} & -0.019^{***} & 0.009        & 0.009        & -0.029^{***} & -0.029^{***} \\
                                     & (0.004)      & (0.004)      & (0.005)      & (0.005)      & (0.006)      & (0.006)      \\
			Duration           & -0.012^{***} & -0.012^{***} & -0.024^{***} & -0.025^{***} & 0.026^{***}  & 0.026^{***}  \\
                                     & (0.003)      & (0.003)      & (0.006)      & (0.006)      & (0.004)      & (0.004)      \\
            Pace                     & 0.002        & 0.002        & 0.013^{*}    & 0.013^{*}    & -0.053^{***} & -0.053^{***} \\
                                     & (0.004)      & (0.004)      & (0.005)      & (0.005)      & (0.006)      & (0.006)      \\
			\addlinespace
			\underline{Political Parties}                                                                                                                                                                                                           \\
			Die Linke                & 0.617^{***}  &              & 0.012        &              & 0.177^{***}  &              \\
                                     & (0.057)      &              & (0.068)      &              & (0.051)      &              \\
			Grüne                    & 0.146^{***}  &              & 0.302^{***}  &              & -0.079^{*}   &              \\
                                     & (0.036)      &              & (0.043)      &              & (0.035)      &              \\
			CDU/CSU                  & -0.249^{***} &              & -0.005       &              & -0.037       &              \\
                                     & (0.041)      &              & (0.049)      &              & (0.039)      &              \\
			FDP                      & -0.124^{*}   &              & -0.079       &              & 0.013        &              \\
                                     & (0.053)      &              & (0.063)      &              & (0.049)      &              \\
			AfD                      & 0.624^{***}  &              & 0.217^{***}  &              & 0.362^{***}  &              \\
                                     & (0.041)      &              & (0.049)      &              & (0.037)      &              \\
			\addlinespace
			\underline{Political Alignment}                                                                                                                                                                                                         \\
			Far-Left                 &              & 0.552^{***}  &              & -0.126       &              & 0.214^{***}  \\
                                     &              & (0.055)      &              & (0.067)      &              & (0.048)      \\
			Center-Right             &              & -0.275^{***} &              & -0.171^{***} &              & 0.016        \\
                                     &              & (0.033)      &              & (0.040)      &              & (0.031)      \\
			Far-Right                &              & 0.557^{***}  &              & 0.078        &              & 0.399^{***}  \\
                                     &              & (0.038)      &              & (0.046)      &              & (0.034)      \\
			\addlinespace
			\underline{Additional Controls}                                                                                                                                                                                                         \\
			User Random Effects   & {Included}                                  & {Included}                                  & {Included}                                     & {Included}           & {Included}           & {Included}    \\
			Monthly Fixed Effects & {Included}                                  & {Included}                                  & {Included}                                     & {Included}           & {Included}           & {Included}     \\
			\midrule
			AIC                   & {\num{-105074}}                              & {\num{-105057}}                              & {\num{-145293}}                                 & {\num{-145250}}       & {\num{-147344}}       & {\num{-147342}} \\
			Observations          & {\num{25246}}                               & {\num{25246}}                               & {\num{25246}}                                  & {\num{25246}}        & {\num{25246}}        & {\num{25246}} \\
			\bottomrule
            \multicolumn{7}{l}{\tiny{Significance levels: $^{***}p<0.001$; $^{**}p<0.01$; $^{*}p<0.05$}} \\
		\end{tabularx}
	}
\end{table}

\end{document}